\documentclass[a4paper,12pt]{article}
\pdfoutput=1 %for pdflatex 
\usepackage{amsmath,amssymb,bm,bbm,mathtools}
\usepackage{ascmac}
\usepackage{cancel}
\usepackage{xcolor} % for pdflatex
\usepackage{graphicx} % for pdflatex
\usepackage{indentfirst}
\usepackage{pdflscape} % landscape for a certain page in pdflatex
\usepackage{colortbl}
\usepackage[bottom]{footmisc}
\usepackage{shadow}
\usepackage{tabularx} 
\usepackage{multirow}
\usepackage{multicol}
\usepackage{marvosym,wasysym}
\usepackage{pifont}
\usepackage{arydshln}
\usepackage{upgreek}
\usepackage{lastpage}
\usepackage{wrapfig}%,wrapfloat}
\usepackage{picinpar}
\usepackage{tabularx}
\usepackage{tikz}
\usepackage{tikz-3dplot} 
\usepackage{pgfplots}
\usetikzlibrary{3d, arrows, arrows.meta, calc, decorations, decorations.markings, intersections, math, matrix, patterns, positioning, shadings, snakes}
\usepackage{circuitikz}
\newdimen\tbaselineshift % for xy.sty and itembox with pdflatex
\usepackage[bookmarks=true,bookmarksnumbered=true,bookmarkstype=toc]{hyperref} 
\hypersetup{
pdfauthor={Tetsuji KIMURA; Shin Sasaki; Kenta Shiozawa},
colorlinks={true},
linkcolor={black},
urlcolor={blue},
filecolor={black},
citecolor={blue}
}
\definecolor{refkey}{rgb}{0.40, 0.55, 0.55}
\definecolor{labelkey}{rgb}{0.40, 0.55, 0.55}
\usepackage{ulem}

\usepackage{nameref}

\usepackage[a4paper,top=33.75truemm,bottom=36.75truemm,left=20truemm,right=20truemm,textwidth=170truemm,textheight=226.5truemm]{geometry}
\makeatletter
\renewcommand{\theequation}{%
   \thesection.\arabic{equation}}
 \@addtoreset{equation}{section}
\makeatother

\definecolor{darkcyan}{HTML}{008B8B}
\makeatletter
\def\tbcaption{\def\@captype{table}\caption}
\def\figcaption{\def\@captype{figure}\caption}
\makeatother

\newcounter{Enumerate}

\DeclareFontFamily{U}{rsf}{}
\DeclareFontShape{U}{rsf}{m}{n}{
  <5> <6> rsfs5 <7> <8> <9> rsfs7 <10-> rsfs10}{}
\DeclareMathAlphabet\Scr{U}{rsf}{m}{n}
\usepackage[mathscr]{eucal}

\newcommand{\del}{\partial}

\newcommand{\half}{\frac{1}{2}}

\newcommand{\LS}{\ \ \ \ \ \ \ \ \ \ }
\newcommand{\ls}{\ \ \ \ \ }
\newcommand{\wt}{\widetilde}

\newcommand{\ve}{\varepsilon}
\newcommand{\ol}{\overline}

\newcommand{\bsubeq}{\begin{subequations}}
\newcommand{\esubeq}{\end{subequations}}

\newcommand{\noi}{\noindent}
\newcommand{\mr}{\mathring}

\newcommand{\nn}{\nonumber}

\newcommand{\vf}{\vphantom{\half}}

\newcommand{\I}{{\rm i}}
\newcommand{\N}{\mathcal{N}}

\renewcommand{\c}{{\sf c}}
\renewcommand{\d}{{\rm d}}
\newcommand{\e}{{\rm e}}

\newcommand{\slb}{\scalebox}

\usepackage{cite}
\begin{document}
\allowdisplaybreaks{
\thispagestyle{empty}
%%%%%%%%% title %%%%%%%%%%%%%
%\begin{flushright}
%since: June 10, 2021 \\
%\timestamp
%\end{flushright}

\vspace*{20mm}

\begin{center}
\noi
\slb{2.2}{
Localized Kaluza-Klein 6-brane
}

\vspace{15mm}

\slb{1.0}{Tetsuji {\sc Kimura}$^{\,{\sf a} \ast}$}, \
\slb{1.0}{Shin {\sc Sasaki}$^{\,{\sf b} \dagger}$} 
\ and \ 
\slb{1.0}{Kenta {\sc Shiozawa}$^{\,{\sf b} \dagger\dagger}$}

\vspace{5mm}

\slb{.85}{
\begin{tabular}{l}
{\renewcommand{\arraystretch}{1.1}
\begin{tabular}{rl}
${\sf a}$ & 
{\sl Center for Physics and Mathematics, Institute for Liberal Arts and Sciences,} 
\\
& {\sl Osaka Electro-Communication University} 
\\
& {\sl 18-8 Hatsucho, Neyagawa, Osaka 572-8530, JAPAN}
\\
& ${\ast}$ {\tt t-kimura \_at\_ osakac.ac.jp}
\end{tabular}
}
\\
\vphantom{A}
\\
{\renewcommand{\arraystretch}{1.1}
\begin{tabular}{rl}
${\sf b}$ & {\sl Department of Physics, Kitasato University}
\\
& {\sl Kitasato 1-15-1, Minami-ku, Sagamihara 252-0373, JAPAN}
\\
& ${\dagger}$ {\tt shin-s \_at\_ kitasato-u.ac.jp}, \ \ 
${\dagger\dagger}$ {\tt k.shiozawa \_at\_ sci.kitasato-u.ac.jp}
\end{tabular}
}
\end{tabular}
}

\end{center}

\vspace{15mm}
%\vfill

%%%%%%%%% abstract %%%%%%%%

\begin{abstract}
We study the membrane wrapping mode corrections to the Kaluza-Klein (KK) 6-brane in eleven dimensions.
We examine the localized KK6-brane in the extended space in $E_{7(7)}$
 exceptional field theory.
In order to discuss the physical origin of the localization in the
extended space, we consider a probe M2-brane in eleven dimensions.
We show that a three-dimensional $\mathcal{N} = 4$ gauge theory is 
naturally interpreted as a membrane generalization of the two-dimensional 
$\mathcal{N} = (4,4)$ gauged linear sigma model for the fundamental string.
We point out that the vector field in the $\mathcal{N}= 4$ model is
 identified as a dual coordinate of the KK6-brane geometry.
We find that the BPS vortex in the gauge theory gives rise to 
the violation of the isometry along the dual direction.
We then show that the vortex corrections are regarded as an
instanton effect in M-theory induced by the probe M2-brane wrapping around the M-circle.

\end{abstract}

%%%%%%%%%%%%%%%%%%%%%%%%%%%%%%%%%%%%%%%%%%%%%%%%%%%%%%%%%%%%%%%%%%%%%%
\newpage
\section{Introduction} \label{sect:introduction}
Dualities are key ingredients to understand the whole picture of string
and M-theories.
In particular, T-duality is one of the most intrinsic properties in string
theory. It arises from the strings wrapped around compactified spaces
and never appears from the viewpoint of particles.
D$p$-branes, which are indispensable objects in string theory, are
related by T-dualities.
The other extended objects such as NS5-branes, F-strings, waves and
exotic branes are also related to the D-branes via S- and T-dualities.
Among other things, the relation between the NS5-branes and the Kaluza-Klein
(KK) 5-branes has been paid much attention \cite{Gregory:1997te}.
The famous Buscher rule \cite{Buscher:1987sk} of T-duality exchanges the NS5-brane smeared
along one transverse direction and the KK5-brane.
The transverse geometry of the KK5-brane is given by the
four-dimensional Taub-NUT space whose isometry comes from the 
smeared direction in the NS5-brane.
This specific T-duality relation gets more involved when one takes into
account the worldsheet instanton effects \cite{Wen:1985jz}.

In \cite{Tong:2002rq}, it was pointed out that the isometry in the
smeared NS5-brane is broken due to the worldsheet instanton effects.
The worldsheet instantons are studied through the examination of
BPS vortices in the two-dimensional $\mathcal{N} = (4,4)$ gauged linear
sigma model (GLSM). In the IR limit, the vortices are interpreted as
worldsheet point-like instantons defined in the string non-linear sigma model.
The violation of the isometry stands for the introduction of the KK massive
modes to the smeared background and the brane is localized in
the compactified direction. This results in the recovery of the original NS5-brane.
Since T-duality exchanges the KK modes and the winding modes, it is natural to
expect that the KK5-brane geometry is modified by the winding modes.
In fact, the worldsheet instanton effects for the KK5-brane geometry
were studied and it was shown that the instantons break the isometry
along the {\it winding direction} \cite{Harvey:2005ab, Jensen:2011jna}.
The resulting geometry depends on the dual winding coordinate
$\tilde{x}$ and the modified background of the KK5-brane
ceases to be a
solution to supergravity.
Although it no longer satisfies the supergravity equations of motion,
it is the genuine T-dual counterpart of the original, non-smeared, NS5-brane.
The same observation holds even for exotic branes \cite{Kimura:2013zva}.

These results lead to difficulties on the interpretation of the modified
---the winding coordinate dependent---
geometries.
However, this problem is naturally solved in the philosophy of double field theory (DFT)
\cite{Hull:2009mi, Siegel:1993xq}.
DFT is a supergravity theory where T-duality symmetry is realized
manifestly in the doubled spacetime $(x^{\mu}, \tilde{x}_{\mu})$.
Since the DFT fields all depend not only on the spacetime coordinate
$x^{\mu}$ but also on the string winding coordinate $\tilde{x}_{\mu}$, 
it is conceivable to incorporate the winding dependent solutions in
the context of DFT.
Indeed, the localized KK5-brane in the winding space is shown to be a
solution to DFT \cite{Berman:2014jsa, Bakhmatov:2016kfn}.
Analogous solutions, such as exotic branes localized in the winding
space were discussed in \cite{Kimura:2018hph}.

The idea of DFT is inherited by exceptional field theory (EFT) 
\cite{Hohm:2013pua, Hohm:2013vpa, Abzalov:2015ega, Musaev:2015ces,
Berman:2015rcc, Bossard:2018utw, Bossard:2019ksx}
where U-duality is realized manifestly (see \cite{Berman:2020tqn} for recent progress and vast applications).
EFT is defined in the extended spaces that are characterized by the
geometric coordinates together with the wrapping coordinates of M-theory
branes. 
With this generalization, it is therefore legitimate to examine the
localized solutions in the extended space of M-theory branes and their
physical origins.
To this end, we focus on the wrapping corrections to the KK6-brane in
eleven dimensions.
This is the natural M-theory uplifting of the winding corrected
KK5-branes in type IIA string theory.

The organization of this paper is as follows.
In the next section, we first introduce the localized KK5-brane in DFT
and discuss its string winding coordinate dependence.
We observe that the localization of the KK5-brane in the winding space is caused by the string worldsheet instanton effects.
We then derive the localized KK6-brane solution in $E_{7(7)}$ exceptional field theory.
We find that the KK6-brane, whose isometry along the dual direction is
broken, is a solution to $E_{7(7)}$ EFT.
In Section \ref{sect:membrane_instantons}, we discuss an
interpretation of the dual coordinate dependence of the localized
KK6-brane. 
We clarify that this is due to instantons caused by the membrane wrapped
around the M-circle.
In Section \ref{sect:3g-gauge},  we introduce the three-dimensional $\mathcal{N} =
4$ Abelian gauge theory and show that this is a natural membrane generalization
of the $\mathcal{N} = (4,4)$ gauged linear sigma model for an F-string propagating on the KK5-brane background.
We write down the corresponding BPS condition in the three-dimensional gauge theory
and find that it is the vortex configuration.
The harmonic function for the deformed solution exhibits the desired
property expected by the instanton effects.
Section \ref{sect:conclusion} is devoted to conclusion and discussions.
The explicit solution of the localized KK6-brane in $E_{7(7)}$ EFT is found in Appendix \ref{sect:localized_KK6}.
Details, conventions and notations of superfields in three dimensions are found in Appendix \ref{sect:SF}.

%%%%%%%%%%%%%%%%%%%%%%%%%%%%%%%%%%%%%%%%%%%%%%%%%%%%%%%%%%%%%%%%%%%%%%%%%
%%%%%%%%%%%%%%%%%%%%%%%%%%%%%%%%%%%%%%%%%%%%%%%%%%%%%%%%%%%%%%%%%%%%%%%%%

\section{KK6-brane in $E_{7(7)}$ exceptional field theory} \label{sect:EFT}

\subsection{Preliminary example -- winding corrections and localized
  KK5-brane}
Before we discuss the localized KK6-brane solution in eleven dimensions,
we first introduce the KK5-brane localized in the string winding space.
This is a solution to DFT.
The dynamical fields of DFT are the generalized metric $\mathcal{H}$ and the
generalized dilation $d$.
They are defined in the $2D$-dimensional doubled space 
characterized by the doubled coordinates $X^M = (x^\mu, \tilde{x}_\mu), \ (\mu = 1, \ldots, D)$.
The DFT action is given by \cite{Hohm:2010pp}:
\begin{align}
S_{\text{DFT}} 
\ &= \ 
\int \d^{2D} X \, \e^{-2d} \, \mathcal{R} (\mathcal{H}, d)
\, , \label{eq:DFT_action}
\end{align}
where the generalized Ricci scalar $\mathcal{R}$ is constructed by $\mathcal{H}$ and
$d$. The action is manifestly invariant under the $O(D,D)$ transformation.
The standard parameterization of the generalized metric $\mathcal{H}$ is
\begin{align}
\mathcal{H}_{MN} 
\ &= \
\left(
\begin{array}{cc}
g^{\mu \nu} & - g^{\mu \rho} B_{\rho \nu} \\
B_{\mu \rho} g^{\rho \nu} & g_{\mu \nu} - B_{\mu \rho} g^{\rho \sigma}
 B_{\sigma \nu}
\end{array}
\right)
\, , \qquad
\e^{-2d} \ = \ \sqrt{-g} \, \e^{-2\phi}
\, , \label{eq:generalized_metric}
\end{align}
where $g_{\mu \nu}, B_{\mu \nu}, \phi$ are the metric, the Kalb-Ramond
$B$-field and the dilation in type II supergravities.
The $D$-dimensional physical spacetime is a slice in the $2D$-dimensional doubled space.
This is defined through the imposition of the strong constraint
$\eta^{MN} \del_M * \del_N * = 0$ where $\eta^{MN}$ is the $O(D,D)$
invariant metric.
Note that any type II supergravity solutions satisfy this constraint.

We focus on the NS5- and the KK5-brane solutions in ten dimensions.
They are incorporated in the following localized DFT monopole ansatz \cite{Berman:2014jsa, Kimura:2018hph}:
\begin{align}
\d s^2_{\mathcal{H}}
\ &= \ 
H (\delta_{ab} - H^{-2} b_{ac} b^c{}_b) \, \d y^a \d y^b
+ H^{-1} \delta^{ab} \, \d \tilde{y}_a \d \tilde{y}_b
+ 2 H^{-1} b_a{}^b \, \d y^a \d \tilde{y}_b
\notag \\
\ & \ \ \ \
+ \eta_{mn} \, \d x^m \d x^n + \eta^{mn} \, \d \tilde{x}_m \d \tilde{x}_n
\, , \notag \\
d
\ &= \ 
\text{const.} - \frac{1}{2} \log H
\, ,
\label{eq:DFT_monopole_ansatz}
\end{align}
where the DFT line element is understood as 
$\d s^2_{\mathcal{H}} = \mathcal{H}_{MN} \d X^M \d X^N$.
Here we label the doubled coordinate as 
$X^M = (x^m,y^a;\tilde{x}_m, \tilde{y}_a), \, (m= 0, \ldots ,5, \, a = 6, \ldots, 9)$.
We define $x^m, y^a$ as the worldvolume and the transverse
directions, respectively. The others $\tilde{x}_m, \tilde{y}_a$ are their 
dual counterparts, namely, the winding coordinates.
The flat metric $\eta_{mn}$ and its inverse $\eta^{mn}$ are defined by
$\eta_{mn} = \mathrm{diag} (-1,\ldots,1)$.
The harmonic function $H$ satisfies $\Box H = 0$ where 
$\textstyle{\Box = \delta^{ab} \frac{\del}{\del y^a} \frac{\del}{\del y^b}}$ 
and 
$b_{ab}$ 
is
a function satisfying 
$3 \del_{[a} b_{bc]} =\varepsilon_{abcd} \del_d H (y)$
Here $\varepsilon_{abcd}$ is the totally antisymmetric symbol.
In the following, we choose specific T-duality frames and extract
physical solutions from \eqref{eq:DFT_monopole_ansatz}.
We start with a solution with 
the geometrical coordinates $x^\mu = (x^m, y^i, y^9)$, $(i = 6,7,8)$. 
From \eqref{eq:DFT_monopole_ansatz} we obtain the
following solution for $g_{\mu \nu}, B_{\mu \nu}, \phi$:
\begin{align}
\d s^2
\ &= \
\eta_{mn} \ \d x^m \d x^n + H \big( \delta_{ij} \, \d y^i \d y^j 
+ (\d y^9)^2 \big)
\, , \notag \\
B
\ &= \ 
\frac{1}{2} (b_{ij} \, \d y^i - 2 \, b_{j9} \, \d y^9) \wedge \d y^j
\, , \notag \\
\e^{2\phi}
\ &= \
H
\, , \label{eq:NS5-brane}
\end{align}
where $\d s^2 = g_{\mu \nu} \d x^{\mu} \d x^{\nu}$,
$\textstyle{H (y^i, y^9) = 1 + \frac{Q}{r^2}}$, $r^2 = (y^i)^2 + (y^9)^2$.
Here $Q$ is a constant.
This is the ordinary NS5-brane solution in type II supergravities
and hence satisfies the section constraint trivially.
We next perform the T-duality transformation of this solution along the $y^9$-direction.
This is implemented by the swap of the coordinates $y^9 \leftrightarrow \tilde{y}_9$. 
After the swapping of the coordinate, we extract the metric from the ansatz \eqref{eq:DFT_monopole_ansatz}.
We then obtain another solution:
\begin{align}
\d s^2
\ &= \
\eta_{mn} \, \d x^m \d x^n + H^{-1} (\d y^9 + b_{i9} \, \d y^i)^2 + H \delta_{ij} \, \d y^i \d y^j
\, , \notag \\
B
\ &= \ 
\frac{1}{2} b_{ij} \, \d y^i \wedge \d y^j
\, , \notag \\
\e^{2\phi}
\ &= \ \text{const.}
\label{eq:localized_KK5}
\end{align}
This is essentially the KK5-brane solution but the non-trivial $B$-field is excited.
In particular, the harmonic function is given by 
$\textstyle{H (y^i, \tilde{y}_9) = 1 + \frac{Q}{r^2}}$, $r^2 = (y^i)^2 + (\tilde{y}_9)^2$.
One notices that the coordinate $\tilde{y}_9$
is not the geometrical one but the T-dualized winding coordinate.
We therefore call the solution \eqref{eq:localized_KK5} the localized KK5-brane
since the brane is localized not only in the ordinary space but also in the winding space.
Such kind of solutions have been discussed in the literature 
\cite{Berkeley:2014nza, Berman:2014jsa,
Bakhmatov:2016kfn,Blair:2016xnn, Lust:2017jox, Berman:2018okd,
Kimura:2018hph}.
Note that the solution \eqref{eq:localized_KK5} 
obtained by the $O(D,D)$ transformation from \eqref{eq:NS5-brane}
also satisfies the $O(D,D)$ invariant strong constraint. 
The physical meaning of the winding coordinate dependence of the
solution becomes apparent when we recover the radius $R_9$ of the compactified
direction and rewrite the harmonic function.
Since the $y^9$- (and $\tilde{y}_9$-) direction is compactified, 
the branes are aligned periodically:
\begin{align}
H (y^i, \tilde{y}_9) 
\ &= \ 
 1 + \sum_{n = - \infty}^{\infty} \frac{Q}{(r')^2 + (\tilde{y}_9 - 2 \pi R_9 n)^2}
\notag \\
\ &= \  
1 + \frac{Q'}{r'} 
\Bigg[
1 + \sum_{n \neq 0} 
\exp 
\left(
\I n \frac{\tilde{y}_9}{R_9} - r' \left| \frac{n}{R_9}\right|
\right)
\Bigg]
\, ,
\label{eq:KK5_harmonic_function}
\end{align}
where we have define a constant $\textstyle{Q' = \frac{Q}{2R_9}}$ and $(r')^2 = (y^6)^2 + (y^7)^2 + (y^8)^2$.
Here the periodic array of the branes along the compactified
direction is evaluated with the Poisson resummation formula,
\begin{align}
\sum^{\infty}_{n = - \infty} f (2 \pi n) 
\ = \ 
\frac{1}{2\pi}
 \sum^{\infty}_{n = - \infty} \int^{\infty}_{- \infty} \d t \, f(t) \e^{\I nt}
\, . \end{align}
It is obvious that $\textstyle{1 + \frac{Q'}{r'}}$ is the harmonic function for the
ordinary KK5-brane of codimension three.
The suggestive form of the $n \neq 0$ corrections in
\eqref{eq:KK5_harmonic_function} makes us evoke instantons.
Indeed, the worldsheet instanton corrections to the KK5-brane geometry was studied in
\cite{Harvey:2005ab}.
In order to study instantons, it is rather useful to start with the two-dimensional
$\mathcal{N} = (4,4)$ gauged linear sigma model (GLSM).
The IR limit of this GLSM is reduced 
to the string worldsheet sigma model whose target
space is the Taub-NUT geometry, namely, the transverse space of the KK5-brane.
GLSM is particularly suitable for evaluating the instanton effects.
In a specific limit of a parameter, it is shown that the two-dimensional GLSM
action is extremized by the BPS vortex configuration.
The corresponding bound of the Euclidean action is given by
\begin{align}
S_n \ = \ 
|n| \zeta - \I n \theta
\, . \label{eq:instanton_action}
\end{align}
Here $n$ is the vortex (instanton) number, $\zeta$ is a Fayet-Iliopoulos (FI) parameter
(cf. \eqref{eq:r1r2r3_vev} in Section \ref{sect:3g-gauge}.)
which is identified with $r'$ of the Taub-NUT space in the IR.
Remarkably, $\theta$ is identified with the dual winding coordinate $\tilde{y}_9$
along the isometry direction.
The standard gauge instanton calculus is available in the
two-dimensional GLSM and the corrections by \eqref{eq:instanton_action}
are encoded into the ones to the harmonic function for the KK5-brane as 
(in our units) \cite{Harvey:2005ab}
\begin{align}
H (r') \ = \ 
1 + \frac{Q'}{r'} \ \ \longrightarrow \ \
H \ = \ 
1 + \frac{Q'}{r'} 
\Bigg[
1 + \sum_{n \neq 0} 
\exp 
\left(
\I n \frac{\tilde{y}_9}{R_9} - r' \left| \frac{n}{R_9}\right|
\right)
\Bigg]
\, .
\end{align}
This precisely agrees with the one in the DFT solution \eqref{eq:KK5_harmonic_function}.
We note that the gauge instantons in GLSM is interpreted as worldsheet instantons
in its IR regime. 
Therefore the origin of the $\tilde{y}_9$-dependence in the localized KK5-brane solution \eqref{eq:localized_KK5}
 is the worldsheet instanton effects.

\subsection{Localized KK6-brane solution in M-theory}
Now we examine an analogous solution that depends on the dual
coordinates in M-theory.
To make contact with the extended space in M-theory, we focus on the
duality relation between the M5-brane and the KK6-brane in eleven dimensions.
The eleven-dimensional spacetime $M_{11}$ is decomposed into $M_4 \times M_7$
where $M_4$ and $M_7$ are the external and the internal spaces.
Since we are interested in the M5- and KK6-branes, we consider the
truncated $E_{7(7)}$ model \cite{Berman:2011jh} where the external 
four-dimensional space $M_4$ is identified with the smeared worldvolume directions of the branes 
and they are essentially ignored. The resulting solutions are all
defined in seven dimensions.
We also ignore all the off-diagonal part between $M_4$ and $M_7$.
The coordinate of the seven-dimensional space $X^{\mu} \, (\mu = 1, \ldots, 7)$ 
together with the M2, M5, KK6 wrapping coordinates 
span the extended space.
The extended space is given by \cite{Coimbra:2011ky}
\begin{align}
T M_7 \oplus \wedge^2 T^* M_7 \oplus \wedge^5 T^*M_7 \oplus (T^*M_7 \otimes \wedge^7 T^*M_7)
\, .
\end{align}
The extended space that consists of $\mathbf{56}$ representation of
$E_{7(7)}$ is characterized by the coordinates
\begin{align}
 \mathbb{X}^M 
\ = \ 
(X^{\mu}, Y_{\mu \nu}, Z^{\mu \nu}, W_{\mu})
\, , \quad 
(M= 1, \ldots, 56)
\, .
\end{align}
The dynamical field of the exceptional field theory is the generalized
metric $\mathcal{M}_{MN}$ which encodes the metric and the 3-form (and
possibly its magnetic dual 6-form) field in eleven-dimensional
supergravity. All the fields and gauge parameters should satisfy the
section constraints \cite{Hohm:2013vpa}
\begin{align}
(t_{\alpha})^{MN} \del_M \del_N * = 0, \quad
(t_{\alpha})^{MN} \del_M * \del_N * = 0, \quad
\Omega^{MN} \del_M * \del_N * = 0.
\label{eq:section_constraints}
\end{align}
Here $(t_{\alpha})^{MN}$ is the generators of $E_{7(7)}$ in the fundamental
representation and 
$\Omega^{MN}$ is the symplectic invariant matrix of $E_{7(7)}$.

A parameterization of the generalized metric $\mathcal{M}_{MN}$ in
$E_{7(7)}$ EFT is derived from the generalized vielbein given in
\cite{Berman:2011jh}.
In our analysis, the 3-form potential $C^{(3)}$ rather than its magnetic
dual 6-form $C^{(6)}$ is relevant.
We therefore keep the metric $g_{\mu \nu}$ and $C^{(3)}$ non-zero and 
do not explicitly use $C^{(6)}$ in the
generalized metric. The result is 
\begin{align}
\d s^2_{\mathcal{M}} 
\ &= \ 
g^{- \frac{1}{2}} 
\Bigg\{
\Bigg[
g_{\mu \nu} + \frac{1}{2} C_{\mu \rho \sigma} g^{\rho \sigma, \lambda
 \tau} C_{\lambda \tau \nu}
+ \frac{1}{32} X_{\mu} {}^{\rho \sigma} g_{\rho \sigma, \alpha \beta}
 X_{\nu} {}^{\alpha \beta} + \frac{1}{24^2} X_{\mu} {}^{\rho \sigma}
 C_{\rho \sigma \alpha} g^{\alpha \beta} X_{\nu} {}^{\lambda \tau}
 C_{\lambda \tau \beta}
\Bigg] \d X^{\mu} \d X^{\nu}
\notag \\
\ & \LS \,
+ 
\Bigg[
g^{\mu \nu, \rho \sigma} + \frac{1}{2} V^{\mu \nu \alpha \beta}
 g_{\alpha \beta, \lambda \tau} V^{\lambda \tau \rho \sigma}
+ \frac{1}{32} X_{\alpha} {}^{\mu \nu} g^{\alpha \beta} X_{\beta}
 {}^{\rho \sigma} 
\Bigg] \d Y_{\mu \nu} \d Y_{\rho \sigma}
\notag \\
\ & \LS \,
+
\Bigg[
g^{-1} g_{\mu \nu, \rho \sigma} + \frac{1}{2} C_{\mu \nu \alpha}
 g^{\alpha \beta} C_{\beta \rho \sigma} 
\Bigg] \d Z^{\mu \nu} \d Z^{\rho \sigma}
+ g^{-1} g^{\mu \nu} \d W_{\mu} \d W_{\nu}
%+
\nn \\
\ & \LS \,
+
\Bigg[
%-
\sqrt{2}
\, C_{\mu \rho \sigma} g^{\rho \sigma, \lambda
 \tau} 
%-
+ \frac{1}{4} X_{\mu} {}^{\rho \sigma} g_{\rho \sigma, \alpha \beta}
 V^{\lambda \tau \alpha \beta} 
%-
+ \frac{1}{48 \sqrt{2}} X_{\mu} {}^{\rho \sigma} C_{\rho \sigma \alpha}
 g^{\alpha \beta} X_{\beta}{}^{\lambda \tau} 
\Bigg] \d X^{\mu} \d Y_{\lambda \tau}
\notag \\
\ & \LS \,
+
\Bigg[
\sqrt{2}
\, g^{-\frac{1}{2}} V^{\mu \nu \alpha \beta}
 g_{\alpha \beta, \rho \sigma} 
+ \frac{1}{4} X_{\alpha} {}^{\mu \nu} g^{\alpha \beta} C_{\beta
 \rho \sigma} 
\Bigg] \d Y_{\mu \nu} \d Z^{\rho \sigma}
\notag \\
\ & \LS \,
+ \sqrt{2}
\, g^{-\frac{1}{2}} C_{\mu \nu \alpha} g^{\alpha
 \rho} \d Z^{\mu \nu} \d W_{\rho}
\notag \\
\ & \LS \,
+ 
\Bigg[
\frac{1}{2 \sqrt{2}} g^{-\frac{1}{2}} X_{\mu} {}^{\alpha \beta}
 g_{\alpha \beta, \lambda \tau}
+
\frac{1}{12 \sqrt{2}} X_{\mu} {}^{\rho \sigma} C_{\rho \sigma \alpha}
 g^{\alpha \beta} C_{\beta \lambda \tau} 
\Bigg] \d X^{\mu} \d Z^{\lambda \tau}
\notag \\
\ & \LS \,
+ \frac{1}{12} g^{- \frac{1}{2}} X_{\mu}{}^{\rho \sigma} C_{\rho \sigma \alpha} g^{\alpha \nu} \, \d X^{\mu} \d W_{\nu}
+ \frac{1}{2 \sqrt{2}} g^{-\frac{1}{2}} X_{\alpha}{}^{\mu \nu} g^{\alpha \rho} \, \d Y_{\mu \nu} \d W_{\rho}
\Bigg\}
\, ,
\label{eq:generalized_metric_parametrization}
\end{align}
In this stage we have performed the sign flipping of the $C$-field from~\cite{Berman:2011jh}.
We also have introduced the following definition:
\begin{align}
g_{\mu \nu, \rho \sigma} 
\ = \ \frac{1}{2} (g_{\mu \rho} g_{\nu \sigma} - g_{\mu \sigma} g_{\nu \rho})
\, , \quad 
V^{\mu \nu \rho \sigma } 
\ = \ 
\frac{1}{3!} \varepsilon^{\mu \nu \rho \sigma \alpha \beta \gamma} C_{\alpha \beta \gamma}
\, , \quad
X_{\mu} {}^{\rho \sigma} 
\ = \ C_{\mu \lambda \tau} V^{\lambda \tau \rho \sigma}
\, .
\end{align}
Here again the line element is understood as 
$\d s^2_{\mathcal{M}} = \mathcal{M}_{MN} \d \mathbb{X}^M \d \mathbb{X}^N$.
As in the case of DFT, we can extract solutions for the metric and the
$C$-field from this parameterization in a fixed U-duality frame.

\paragraph{From KK6-brane to smeared M5-brane:}
We first start with the ordinary KK6-brane and see how this is dualized to
the M5-brane. 
The eleven-dimensional spacetime coordinate $x^{\mu}$ is labeled by 
$x^{\mu} = (t, x^a, y^i, z), \, (a = 1, \ldots, 6, \, i = 1, \ldots,
3)$.
Here $(t, x^a)$ and $(y^i,z)$ are identified with the worldvolume and the
transverse directions of the KK6-brane, respectively.
We choose the seven-dimensional internal space as $X^{\mu} = (u, v, w, y^i, z)$
and $t, x^{3,4,5}$ are the external reduced directions.
Here we denote $x^1 = u, x^2 = v, x^6 = w$.
The ordinary KK6-brane of codimension three in seven dimensions is given by
\begin{align}
\d s^2 
\ = \ 
\d u^2 + \d v^2 + \d w^2  
+ H^{-1} \left[ \d z + A_i \, \d y^i \right]^2 + H \, \d \vec{y}^{\,2}
\, , \label{eq:KK6cod3}
\end{align}
where the KK vector $A_i$ satisfies
$\textstyle{\del_{[i} A_{j]} = \frac{1}{2} \varepsilon_{ijk} \del_k H}$ 
and the harmonic function is defined by
$\textstyle{H = 1 + \frac{h}{|\vec{y}|}}$. 
Here $h$ is a constant.
This background metric is embedded into the generalized metric $\mathcal{M}_{MN}$ 
by using the parameterization \eqref{eq:generalized_metric_parametrization} 
(see the equation (3.21) in \cite{Berman:2014jsa}
\footnote{
Note that the parameterization of the generalized metric in \cite{Berman:2014jsa}
is slightly different from \eqref{eq:generalized_metric_parametrization}.
This results in the different scaling factors discussed in \cite{Berman:2014jsa}
as we will see in due course.
}
).
Since \eqref{eq:KK6cod3} is a solution to ordinary supergravity,
it trivially satisfies the section constraints \eqref{eq:section_constraints}.
We stress that $X_{\mu} {}^{\rho \sigma}$-dependent parts all vanish for the solution of
codimension three.
This drastically simplify the calculations.
The duality transformation from the KK6-brane to the M5-brane is 
implemented by the swaps of the extended coordinates together with
appropriate scaling transformations \cite{Berman:2014jsa}.
The swaps of the coordinates are given by
\begin{alignat}{2}
X^z \ &\leftrightarrow \ - Y_{wz}
& \LS
W_z \ &\leftrightarrow \  Z^{wz}
\notag \\
X^w \ &\leftrightarrow \  Y_{uz}
& \LS
W_w \ &\leftrightarrow \  Z^{uz}
\notag \\
Y_{uv} \ &\leftrightarrow \  - Y_{vz}
& \LS
Z^{uv} \ &\leftrightarrow \  Z^{vz}
\notag \\
Y_{ui} \ &\leftrightarrow \  - Y_{iz}
& \LS
Z^{ui} \ &\leftrightarrow \  Z^{iz}
\notag \\
Y_{vi} \ &\leftrightarrow \  \frac{1}{2} \varepsilon_{ijk} Z^{jk}
& \LS
Z^{vi} \ &\leftrightarrow \ \frac{1}{2} \varepsilon^{ijk} Y_{jk}
\, .
\label{eq:KK6toM5}
\end{alignat}
The dualized metric and the $C$-field are extracted from the following ansatz:
\begin{align}
\d s^2_{\mathcal{M}} 
\ &= \ 
g^{- \frac{1}{2}} 
\Bigg\{
\Bigg[
g_{\mu \nu} 
+ \frac{1}{2} 
\e^{2 \gamma_1}
C_{\mu \rho \sigma} g^{\rho \sigma, \lambda \tau} C_{\lambda \tau \nu}
\Bigg] \d X^{\mu} \d X^{\nu}
\notag \\
\ & \LS \,
+ 
\Bigg[
\e^{2 \alpha_1}
g^{\mu \nu, \rho \sigma} 
+ 
\frac{1}{2} 
\e^{2 \gamma_2}
V^{\mu \nu \alpha \beta}
 g_{\alpha \beta, \lambda \tau} V^{\lambda \tau \rho \sigma}
\Bigg] \d Y_{\mu \nu} \d Y_{\rho \sigma}
\notag \\
\ & \LS \,
+
\Bigg[
\e^{2 \alpha_2}
g^{-1} g_{\mu \nu, \rho \sigma} 
+ \frac{1}{2} \e^{2 \gamma_3} C_{\mu \nu \alpha} g^{\alpha \beta} C_{\beta \rho \sigma} 
\Bigg] \d Z^{\mu \nu} \d Z^{\rho \sigma}
\notag \\
\ & \LS \,
+ \e^{2 \alpha_3} g^{-1} g^{\mu \nu} \, \d W_{\mu} \d W_{\nu}
+ \sqrt{2}
\, \e^{2 \beta_1} C_{\mu \rho \sigma} g^{\rho \sigma, \lambda \tau} \, \d X^{\mu} \d Y_{\lambda \tau}
\notag \\
\ & \LS \,
+ \sqrt{2}
\, \e^{2 \beta_2} g^{-\frac{1}{2}} V^{\mu \nu \alpha \beta} g_{\alpha \beta, \rho \sigma} \, \d Y_{\mu \nu} \d Z^{\rho \sigma}
+ \sqrt{2}
\, \e^{2 \beta_3} g^{-\frac{1}{2}} C_{\mu \nu \alpha} g^{\alpha \rho} \, \d Z^{\mu \nu} \d W_{\rho}
\Bigg\}
\, .
\end{align}
Here $\alpha_i, \beta_i, \gamma_i \, (i=1,2,3)$ are the scale parameters.
After the swapping of the coordinates \eqref{eq:KK6toM5}, the explicit form of the components of $g_{\mu \nu}$
and $C_{\mu \nu \rho}$ together with the scaling factors are determined
consistently.
The result is
\begin{alignat}{3}
\e^{2 \alpha_1} 
\ &= \ 
\frac{1}{2} H^{-\frac{4}{5}}
\, , & \ls
\e^{2 \beta_1} 
\ &= \ 
\frac{1}{\sqrt{2}} H^{-\frac{4}{5}}
\, , & \ls
\e^{2 \gamma_1} 
\ &= \ 
H^{-\frac{4}{5}}
\, , \notag \\
\e^{2 \alpha_2} 
\ &= \ 
\frac{1}{2} H^{-\frac{8}{5}}
\, , & \ls
\e^{2 \beta_2} 
\ &= \ 
\frac{1}{2 \sqrt{2}} H^{-2}
\ , & \ls
\e^{2 \gamma_2} 
\ &= \ \frac{1}{4} H^{-\frac{12}{5}}
\, , \notag \\
\e^{2 \alpha_3} 
\ &= \ H^{-\frac{12}{5}}
, & \ls
\e^{2 \beta_3} 
\ &= \ 
\frac{1}{\sqrt{2}} H^{-\frac{14}{5}}
\, , & \ls
\e^{2 \gamma_3} 
\ &= \ 
\frac{1}{2} H^{-\frac{16}{5}}
\, , \label{eq:scale1}
\end{alignat}
and
\begin{gather}
\d s^2 
\ = \ 
H^{- \frac{3}{5}} \big[ \d u^2 + \d v^2 \big]
+ H^{\frac{2}{5}} \big[ \d w^2 + \d \vec{y}^{\, 2} + \d z^2 \big]
\, , \notag \\
C_{izw} \ = \ A_i
\, , \ls
H \ = \ 1 + \frac{h}{r}
\, , \ls
r^2 \ = \ \vec{y}^{\, 2}
\, .
\end{gather}
This is the M5-brane dimensionally reduced to seven dimensions.
Note that all the solutions are given in the Einstein frame.
We stress that the resulting solution has non-zero components of the $C$-field only in 
$C_{i z w}$. This is the direct consequence of the M5-brane of
codimension three -- the M5-brane smeared along the two transverse directions.
As we will see in the following, when we consider the M5-brane of higher codimensions,
there are more non-zero components of the $C$-field which inevitably changes the 
structure of the generalized metric.

\paragraph{Localized KK6-brane from M5-brane of codimension four:}
We next examine a backward transformation from the M5-brane to the KK6-brane but with the 
assumption of codimension four solution.
This will provide the direct M-theory uplifting of the localized KK5-brane in DFT.

The M5-brane of codimension four dimensionally reduced to seven dimensions 
is apparently a solution to supergravity:
\begin{align}
\d s^2 
\ &= \ 
H^{-\frac{3}{5}} (\d u^2 + \d v^2) + H^{\frac{2}{5}} (\d w^2 + \d \vec{y}^{\, 2} + \d z^2)
\, , \notag \\
C 
\ &= \ 
\frac{1}{2!} C_{ijw} \, \d y^i \wedge \d y^j \wedge \d w 
+ C_{izw} \, \d y^i \wedge \d z \wedge \d w
\, , \notag \\
H 
\ &= \ 
1 + \frac{h}{\vec{y}^{\, 2} + z^2}
\, . \label{eq:M5cod4}
\end{align}
This is obtained by the dimensional reduction of the M5-brane solution
in eleven-dimensional supergravity to seven dimensions and by moving to
the Einstein frame.
Since the $C$-field is determined by the relation
\vspace{-2mm}
\begin{align}
4 \del_{[a} C_{bcd]} \ = \ \varepsilon_{abcde} \del_e H
\, ,
\end{align}
there are more non-zero components than $C_{i z w}$ for the $C$-field.
This results in $X_{\mu} {}^{\rho \sigma} \neq 0$ which distinguishes
the analysis from that of codimension three and 
the embedding of the solution \eqref{eq:M5cod4} into the
generalized metric is treated with care.
We have
\begin{align}
\d s^2_{\mathcal{M}} 
\ &= \
g^{- \frac{1}{2}} 
\Bigg\{
\Bigg[
g_{\mu \nu} 
+ \frac{1}{2} \e^{2 \gamma_1} C_{\mu \rho \sigma} g^{\rho \sigma, \lambda \tau} C_{\lambda \tau \nu}
+ \frac{1}{32} \e^{2 \delta_1} X_{\mu}{}^{\rho \sigma} g_{\rho \sigma, \alpha \beta} X_{\nu}{}^{\alpha \beta} 
\notag \\
\ & \LS \LS 
+ \frac{1}{24^2} \e^{2 \delta_2} X_{\mu}{}^{\rho \sigma} C_{\rho \sigma \alpha} g^{\alpha \beta} X_{\nu}{}^{\lambda \tau} C_{\lambda \tau \beta}
\Bigg] \d X^{\mu} \d X^{\nu}
\notag \\
\ & \LS \,
+ 
\Bigg[
\e^{2 \alpha_1} g^{\mu \nu, \rho \sigma} 
+ 
\frac{1}{2} \e^{2 \gamma_2} V^{\mu \nu \alpha \beta} g_{\alpha \beta, \lambda \tau} V^{\lambda \tau \rho \sigma}
+ \frac{1}{32} \e^{2 \delta_3} X_{\alpha}{}^{\mu \nu} g^{\alpha \beta} X_{\beta}{}^{\rho \sigma} 
\Bigg] \d Y_{\mu \nu} \d Y_{\rho \sigma}
\notag \\
\ & \LS \,
+
\Bigg[
\e^{2 \alpha_2} g^{-1} g_{\mu \nu, \rho \sigma} 
+ \frac{1}{2} \e^{2 \gamma_3} C_{\mu \nu \alpha} g^{\alpha \beta} C_{\beta \rho \sigma} 
\Bigg] \d Z^{\mu \nu} \d Z^{\rho \sigma}
+ \e^{2 \alpha_3} g^{-1} g^{\mu \nu} \, \d W_{\mu} \d W_{\nu}
\notag \\
\ & \LS \,
+
\Bigg[
 \sqrt{2}
\, \e^{2 \beta_1} C_{\mu \rho \sigma} g^{\rho \sigma, \lambda \tau} 
+ \frac{1}{4} \e^{2 \delta_4} X_{\mu}{}^{\rho \sigma} g_{\rho \sigma, \alpha \beta} V^{\lambda \tau \alpha \beta} 
+ \frac{1}{48 \sqrt{2}} \e^{2 \delta_5} X_{\mu}{}^{\rho \sigma} C_{\rho \sigma \alpha} g^{\alpha \beta} X_{\beta}{}^{\lambda \tau} 
\Bigg] \d X^{\mu} \d Y_{\lambda \tau}
\notag \\
\ & \LS \,
+
\Bigg[
 \sqrt{2}
\, \e^{2 \beta_2} g^{-\frac{1}{2}} V^{\mu \nu \alpha \beta} g_{\alpha \beta, \rho \sigma} 
+ \frac{1}{4} \e^{2 \delta_9} X_{\alpha}{}^{\mu \nu} g^{\alpha \beta} C_{\beta \rho \sigma} 
\Bigg] \d Y_{\mu \nu} \d Z^{\rho \sigma}
\notag \\
\ & \LS \,
+ \sqrt{2}
\, \e^{2 \beta_3} g^{-\frac{1}{2}} C_{\mu \nu \alpha} g^{\alpha \rho} \, \d Z^{\mu \nu} \d W_{\rho}
\notag \\
\ & \LS \,
+ 
\Bigg[
\frac{1}{2 \sqrt{2}} g^{-\frac{1}{2}} \e^{2 \delta_6} X_{\mu}{}^{\alpha \beta} g_{\alpha \beta, \lambda \tau}
+
\frac{1}{12 \sqrt{2}} \e^{2 \delta_7} X_{\mu}{}^{\rho \sigma} C_{\rho \sigma \alpha} g^{\alpha \beta} C_{\beta \lambda \tau} 
\Bigg] \d X^{\mu} \d Z^{\lambda \tau}
\notag \\
\ & \LS \,
+ \frac{1}{12} \e^{2 \delta_8} g^{- \frac{1}{2}} X_{\mu}{}^{\rho \sigma} C_{\rho \sigma \alpha} g^{\alpha \nu} \, \d X^{\mu} \d W_{\nu}
+ \frac{1}{2 \sqrt{2}} \e^{2 \delta_{10} } g^{-\frac{1}{2}} X_{\alpha}{}^{\mu \nu} g^{\alpha \rho} \, \d Y_{\mu \nu} \d W_{\rho}
\Bigg\}
\, ,
\label{eq:rescaled_gen_metric}
\end{align}
where we have introduced the scaling factors $\alpha_i, \beta_i,
\gamma_i, \, (i=1,2,3)$ determined in \eqref{eq:scale1}.
Unlike the case of the codimension three, we also introduce the new scaling factors $\delta_i \, (i=1,\ldots,10)$
in the $X_{\mu}{}^{\rho \sigma}$ sector for later convenience. They are 
\begin{alignat}{3}
\e^{2 \delta_1} 
\ &= \ H^{-\frac{12}{5}}
\, , & \ls
\e^{2 \delta_3} 
\ &= \ \frac{1}{2} H^{-4}
\, , & \ls 
\e^{2 \delta_4} 
\ &= \  \frac{1}{2} H^{-\frac{12}{5}}
\, , \notag \\
\e^{2 \delta_6} 
\ &= \ 
\frac{1}{\sqrt{2}} H^{-2}
\, , & \ls 
\e^{2 \delta_9} 
\ &= \  \frac{1}{2} H^{-\frac{16}{5}}
\, , & \ls
\e^{2 \delta_{10}} 
\ &= \ 
\frac{1}{\sqrt{2}} H^{-\frac{14}{5}}
\, ,
\label{eq:scale2}
\end{alignat}
and $\e^{2 \delta_2} = \e^{2 \delta_5} = \e^{2 \delta_5} = \e^{2 \delta_7} = \e^{2 \delta_8} = 1$.

Now we perform the backward duality transformation.
This is achieved again by the swaps of the coordinates \eqref{eq:KK6toM5}
and the scale transformation.
After the coordinate swaps, the metric and the $C$-field are extracted 
from the same form of the generalized metric
\eqref{eq:rescaled_gen_metric} but with the new scaling factors
$\alpha'_i, \beta'_i, \gamma'_i, \delta'_i$.
We find that they are consistently determined to be
(see Appendix for details)
\begin{alignat}{3}
\e^{2 \alpha'_1} 
\ &= \ 
\frac{1}{2}
\, , & \ls
\e^{2 \beta'_1} 
\ &= \ 
 \frac{1}{\sqrt{2}}
\, , & \ls 
\e^{2 \gamma'_1} 
\ &= \ 1
\, , \notag \\
\e^{2 \alpha'_2} 
\ &= \ 
\frac{1}{2}
\, , & \ls 
\e^{2 \beta_2'} 
\ &= \ 
 \frac{1}{2 \sqrt{2}} H^{-1}
\, , & \ls
\e^{2 \gamma'_2} 
\ &= \ 
\frac{1}{4} H^{-2}
\, , \notag \\
\e^{2 \alpha'_3} 
\ &= \ 1
\, , & \ls
\e^{2 \beta'_3} 
\ &= \ 
 \frac{1}{\sqrt{2}} H^{-1}
\, , & \ls 
\e^{2 \gamma'_3} 
\ &= \ 
\frac{1}{2} H^{-2}
\, .
\label{eq:scale3}
\end{alignat}
Then the KK6-brane of codimension four is given by
\begin{align}
\d s^2 
\ &= \ 
\d u^2 + \d v^2 + \d w^2 + H \d \vec{y}^{\, 2} + H^{-1} (\d z + C_{izw} \, \d y^i)^2
\, , \notag \\
C 
\ &= \ 
\frac{1}{2!} C_{ijw} \, \d y^i \wedge \d y^j \wedge \d w
\, . \label{eq:localized_KK6}
\end{align}
We find that $X_{\mu}{}^{\rho \sigma} = 0$ for this metric and
the $C$-field and terms with the scaling factors $\delta'_i$ all vanish.
Most notably, the harmonic function in the solution \eqref{eq:localized_KK6} 
is given by
\begin{align}
H \ = \ 1 + \frac{h}{\vec{y}^{\, 2} + Y_{wz}^2}
\, . \label{eq:KK6_harmonic_function}
\end{align}
Since the geometrical coordinates are $(t, x^a, y^i, z)$ in this frame, 
the solution depends on the dual (wrapping) coordinate $Y_{wz}$ as
expected.
Once the $Y_{wz}$ direction is smeared, this solution reduces to the ordinary KK6-brane.
When one takes the $w$-direction as the M-circle and relabel $B_{ij} =
C_{ijw}$, $Y_{wz} = \tilde{y}_9$, the solution \eqref{eq:localized_KK6} reduces to the localized
KK5-brane discussed in Section \ref{sect:EFT} (Fig.\ref{fig:KK5-KK6}).
Hence the solution \eqref{eq:localized_KK6} is the M-theory analogue of the localized KK5-brane \eqref{eq:localized_KK5}.
We call \eqref{eq:localized_KK6} the localized KK6-brane solution.
%%%%%%%
\begin{figure}
\begin{center}
\includegraphics[scale=0.45]{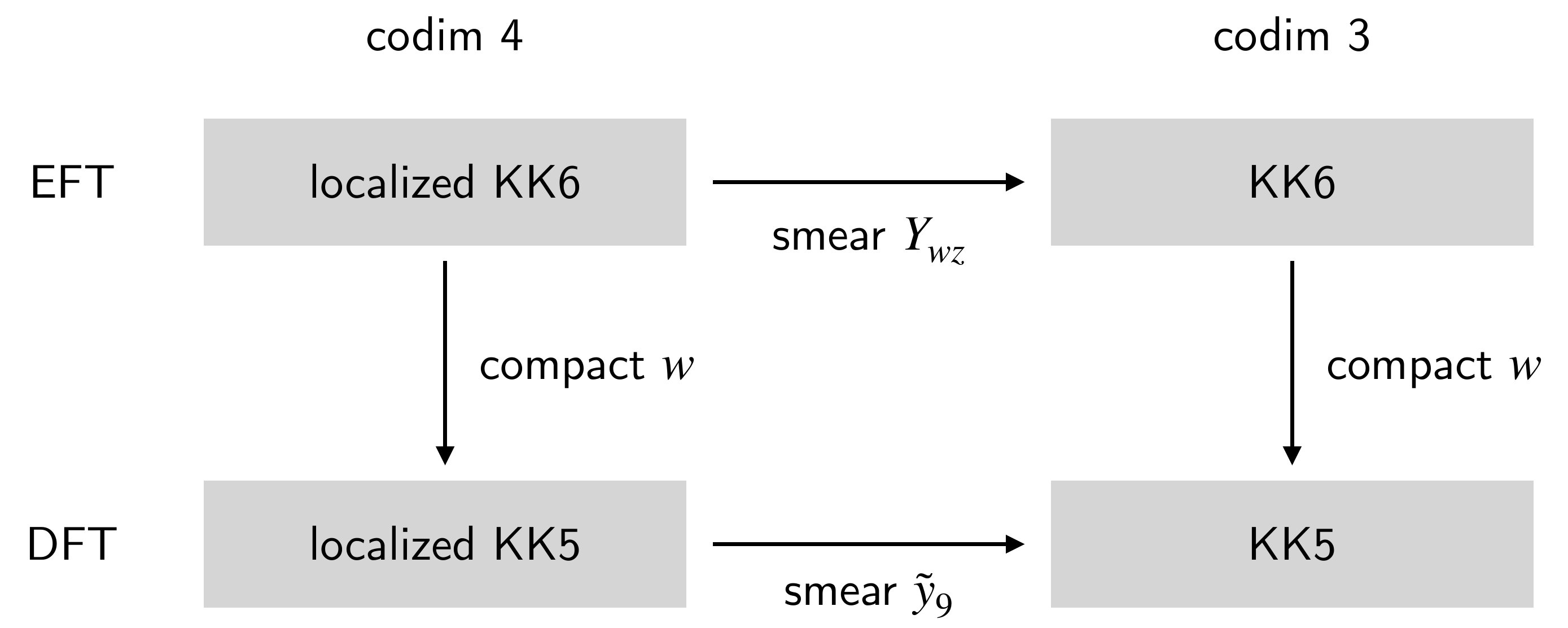}
\caption{The localized and the ordinary KK5- and KK6-branes in DFT and EFT.}
\label{fig:KK5-KK6}
\end{center}
\end{figure}
%%%%%%%

When we recover the radius $R$ of the compactified direction and rewrite
the harmonic function \eqref{eq:KK6_harmonic_function}, we have
\begin{align}
H \ = \ 
1 + \frac{h'}{|\vec{y}|} 
\left[
1 + \sum_{n \neq 0} 
\exp
\left(
\I n \frac{Y_{wz}}{R} - |\vec{y}| \left| \frac{n}{R} \right|
\right)
\right]
\, , \label{eq:KK6_harmonic_function_corrections}
\end{align}
where $\textstyle{h'=\frac{h}{2R}}$.
Once again, $\textstyle{1 + \frac{h'}{|\vec{y}|}}$ part is the harmonic function for
the ordinary KK6-brane of codimension three \eqref{eq:KK6cod3}.
The $n \neq 0$ corrections strongly suggest the existence of instantons.
In order to reveal the origin of the ``instantons'' appear in \eqref{eq:KK6_harmonic_function_corrections}, 
we next discuss the geometry of the KK6-brane.

\section{Membrane and worldsheet instantons in KK6 and KK5 geometries} \label{sect:membrane_instantons}

In this section, we examine the instanton effects in eleven dimensions
and discuss the origin of the wrapping corrections to the KK6-brane geometry.

As we have mentioned in the introduction, the winding corrections to the
KK5-brane geometry in type IIA string theory is induced by the string
worldsheet instantons.
% \cite{Wen:1985jz}.
The worldsheet instantons are defined as a map $\Sigma_2 \to \gamma_2$ where $\Sigma_2$ and $\gamma_2$ are the 
Euclidean worldsheet and a two-cycle in spacetime, respectively.
We consider $S^2$ as the topology of $\Sigma_2$.
The transverse geometry of the multiple KK5-branes in ten dimensions is the
 multi-centered Taub-NUT space \cite{Okuyama:2005gx, Kimura:2018hph}, where an $S^2$ appears between the centers. 
Then the worldsheet instantons are classified by the homotopy group
$\pi_2 (S^2) = \mathbb{Z}$.
Since the KK5-brane is obtained by the dimensional reduction of the
KK6-brane in eleven dimensions, 
a natural candidate of the M-theory origin of the winding corrections is
the membrane instantons \cite{Becker:1995kb, Harvey:1999as}.
The membrane instantons are defined by a map $\Sigma_3 \to \gamma_3$ where $\Sigma_3 = S^3$ is the 
Euclidean M2-brane worldvolume and $\gamma_3$ is a three-cycle in eleven-dimensional spacetime.
However, since the background geometry of the multiple KK6-brane is again the multi-centered Taub-NUT space, 
 there are no three-cycles in the geometry and the Euclidean M2-brane seems not to wrap on the geometry.
This apparent puzzle can be reconciled as follows.

We consider the brane configuration in Table \ref{tb:KK6_KK5}.
%%%%%
\begin{table}
\begin{center}
\begin{tabular}{c|c|c|c|c|c|c|c|c|c|c|c}
    & 0 & 1 & 2 & 3 & 4 & 5 & 6 & 7 & 8 & 9 & M \\
\hline
KK6 & $\times$ & $\times$ & $\times$ & $\times$ & $\times$ & $\times$ &
			     &  &  & $\circ$ & $\times$ \\
\hline
M2  & & & & & & & & & $\times$ & $\times$ & $\times$ \\
\hline
cycles     & & & & & & & &  &  \multicolumn{2}{c|}{$S^2$}  & $S^1$ \\
\multicolumn{12}{c}{}
\\
\hdashline
IIA KK5 & $\times$ & $\times$ & $\times$ & $\times$ & $\times$ & $\times$ &
			     &  &  & $\circ$ &  \\
\hline
IIA F1  & & & & & & & & & $\times$ & $\times$ & 
\end{tabular}
\end{center}
\caption{
The configuration of Euclidean M2-brane wrapping on the M-circle $S^1$ and $S^2$ in the KK6 geometry.
The worldvolume and the isometry directions are denoted as $\times$ and
 $\circ$, respectively.
The lower table is the corresponding configuration in ten-dimensional type IIA theory.
}
\label{tb:KK6_KK5}
\end{table}
%%%%%
One finds that the Euclidean M2-brane can wrap on the M-circle $S^1$ and
the two-sphere $S^2$ in the 
%KK6 
Taub-NUT geometry.
Indeed, the homology group for the three-cycles is calculated as 
\begin{align}
H_3 (S^1 \times S^2) 
\ = \ 
\mathbb{Z}
\, .
\end{align}
Then the membrane instantons are classified by the homotopy group
\begin{align}
\pi_3 (S^1 \times S^2) 
\ \simeq \ 
\pi_3 (S^1) \oplus \pi_3 (S^2) 
\ = \ 
\{ e \} \oplus \mathbb{Z} 
\ = \ 
\mathbb{Z}
\, .
\end{align}
Therefore by the introduction of the M-circle, the Euclidean M2-brane
can wrap on the three-cycle and the KK6 geometry supports the membrane
instantons.
This is a natural M-theory origin of the worldsheet instanton corrections to the KK5-brane.
Schematically, this is sketched out as 
\begin{align}
\text{(membrane instanton)} \sim& \
\exp
\left[
- \frac{\mathrm{vol} (\Sigma_3)}{l_{11}^3}
\right]
= 
\exp
\left[
- \frac{\mathrm{vol} (S^2)}{\alpha'}
\right]
\sim
\text{(worldsheet instanton)},
\end{align}
where we have used the relation $l_{11} = g_s^{\frac{1}{3}} l_s$
among the eleven-dimensional Planck length $l_{11}$, 
the string coupling constant $g_s$ and the string length $l_s$.
The M-circle radius is given by $R_{\mathrm{M}} = g_s l_s$.
The analogous mechanism on the moduli spaces for the vector and the
hypermultiplets in four and five dimensions was studied \cite{Becker:1995kb}.

When there is only a single KK6-brane, the geometry involves an open cigar and there are no two-cycles anymore.
However we find a kind of instantons in this geometry.
We regard the closed membrane worldvolume $S^3$ as the Hopf fibration $S^1$ over $S^2$.
The $S^1$ is identified with the M-circle and the remaining $S^2$ is
decomposed into a disk and an infinity point: $S^2 = D^2_{\Sigma} \cup \{ \infty_{\Sigma} \}$.
Then we find out a map:
\begin{align}
D^2_{\Sigma} \to D^2
\, , \quad 
\{ \infty_{\Sigma} \} \to \{ \infty \}
\, ,
\end{align}
where $D^2$ is a disk defined by the open cigar in the KK6 geometry and $\{\infty \}$ is the infinity point in the spacetime.
This is an analogy of the disk instantons for the worldsheet
\cite{Okuyama:2005gx, Kimura:2018hph}.

In order to specify the field configuration that corresponds to the
wrapped membrane instantons, we next introduce a three-dimensional
gauge theory which is a membrane generalization of the
two-dimensional 
GLSM discussed in \cite{Tong:2002rq, Harvey:2005ab}.

%%%%%%%%%%%%%%%%%%%%%%%%%%%%%%%%%%%%%%%%%%%%%%%%%%%%%%%%%%%%%%%%%%%%%%
%%%%%%%%%%%%%%%%%%%%%%%%%%%%%%%%%%%%%%%%%%%%%%%%%%%%%%%%%%%%%%%%%%%%%%

\section{Three-dimensional gauge theory}
\label{sect:3g-gauge}

In this section we study
three-dimensional $\N=4$ gauge theory whose low energy theory could be interpreted as the worldvolume sigma model whose target space is the Taub-NUT space. 
This is also an explicit description of an Abelian gauge theory studied in \cite{Kapustin:1999ha}.
Analyzing its quantum corrections by instanton effects, we could discuss the physical meaning of the corrections {in EFT in the previous sections}.

%%%%%%%%%%%%%%%%%%%%%%%%%%%%%%%%%
\subsection{Lagrangian}
\label{sect:3d-L}

First, we describe a three-dimensional $\N=4$ Abelian gauge theory. 
This is easily discussed by uplifting the two-dimensional $\N=(4,4)$ GLSM for a KK5-brane in the superfield formalism \cite{Tong:2002rq, Harvey:2005ab, Okuyama:2005gx, Kimura:2013fda}\footnote{Compared with these references, we can change the relative signs in the third line via flipping the signs of $\Phi$, $s$, $\Psi$, and $\wt{Q} Q$ appropriately. We can also extend the model to $U(1)^N$ gauge theory.}:
\begin{align}
\mathcal{L}_{\text{3d}}
\ &= \ 
\frac{1}{e^2} \int \d^4 \theta \, \Big\{ - \Sigma^2 + |\Phi|^2 \Big\}
+ \int \d^4 \theta \, \Big\{ 
\frac{1}{g^2} |\Psi|^2
+ \frac{g^2}{2} \big( \Gamma + \ol{\Gamma} + 2 \c V \big)^2
\Big\}
\nn \\
\ & \ \ \ \ 
+ \int \d^4 \theta \, \Big\{
|Q|^2 \, \e^{+ 2 V} + |\wt{Q}|^2 \, \e^{- 2 V}
\Big\}
\nn \\
\ & \ \ \ \ 
+ \Big\{ 
\sqrt{2} \int \d^2 \theta \, \Big( \wt{Q} \Phi Q - \c \, (s - \Psi) \Phi \Big)
+ \text{(h.c.)}
\Big\}
+ 2 \sqrt{2} \, \c \int \d^4 \theta \, t^3 \, V
\nn \\
\ & \ \ \ \ 
- \sqrt{2} \, \c \, {\ve}^{mnp} \, \del_m (B_n A_{p})
\, . \label{3d-GLSMKK}
\end{align}
Let us explain various parameters and fields in this Lagrangian.
\begin{itemize}
\item $e$ is the gauge coupling constant whose mass dimension is $1/2$. 
$g$ is the dimensionless sigma model coupling.
$\c$ is a constant whose dimension is $1/2$. 
This is reduced to a dimensionless constant when we perform the dimensional reduction of this three-dimensional theory to two-dimensional $\N=(4,4)$ GLSM.

\item The $\N=4$ vector multiplet consists of an $\N=2$ real linear
      superfield $\Sigma$ and an $\N=2$ chiral superfield
      $\Phi$\footnote{The definition and the component expansion
 of superfields are exhibited in appendix \ref{sect:SF}.}.
The pair $( \Psi, \Gamma )$ builds an $\N=4$ neutral hypermultiplet.
Notice that the chiral superfield $\Gamma$ is coupled to the vector superfield $V$ via the St\"{u}ckelberg coupling.
The pair of two $\N=2$ charged chiral superfields $( Q, \wt{Q} )$ builds an $\N=4$ charged hypermultiplet.

\item The parameter $s$ in the third line is complex, while $t$ is real. They are the Fayet-Iliopoulos (FI) parameters. They behave as a triplet under the $SU(2)$ R-symmetry rotation\footnote{More precisely, the $SU(2)_H$ R-symmetry in the Higgs branch which rotates three real scalar fields $(r^1, r^2, -g^2 \wt{r}^3)$ in the $\N=4$ neutral hypermultiplet $( \Psi, \Gamma )$.}.

\item The term in the fourth line is an uplifted version of $\ve^{mn} \del_m (\theta A_n)$ in two-dimensional theory \cite{Harvey:2005ab}, 
which plays a central role in analyzing the winding corrections in the Taub-NUT space \cite{Harvey:2005ab}. 
A vector field $B_m$ is the uplifted field of the periodic scalar $\theta$ in the two-dimensional GLSM, where $\theta$ represents ``winding coordinate'' in the Taub-NUT space.
Note that ${\ve}^{mnp}$ is the totally antisymmetric symbol whose normalization is ${\ve}^{012} = +1$.
\end{itemize}

Expanding all the superfields and integrating their auxiliary fields out, we obtain the interacting Lagrangian given by component fields:
\begin{align}
\mathcal{L}
\ &= \ 
- \frac{1}{e^2} \Big\{ 
\frac{1}{4} {F}_{mn} {F}^{mn}
+ \frac{1}{2} (\del_m \sigma)^2
+ |\del_m \phi|^2
\Big\}
- \sqrt{2} \, \c \, {\ve}^{mnp} \, \del_m (B_n A_{p})
\nn \\
%%%
\ & \ \ \ \ 
- \frac{1}{2 g^2} \Big\{ (\del_m r^1)^2 + (\del_m r^2)^2 + (\del_m (- g^2 \wt{r}^3))^2 \Big\}
- \frac{g^2}{2} \big( \del_m \wt{r}^4 + \sqrt{2} \, \c \, A_m \big)^2
\nn \\
%%%
\ & \ \ \ \ 
+ \frac{1}{e^2} \Big\{
\I (\del_m \ol{\lambda} \gamma^m \lambda)
+ \I (\del_m \ol{\wt{\lambda}} \gamma^m \wt{\lambda})
\Big\}
+ \frac{\I}{g^2} (\del_m \ol{\chi} \gamma^m \chi)
+ \I g^2 (\del_m \ol{\wt{\chi}} \gamma^m \wt{\chi}) 
\nn \\
%%%
\ & \ \ \ \ 
- |D_m q|^2 - |D_m \wt{q}|^2 
+ \I (D_m \ol{\psi} \gamma^m \psi)
+ \I (D_m \ol{\wt{\psi}} \gamma^m \wt{\psi})
\nn \\
%%%
\ & \ \ \ \ 
- \big( \sigma^2 + 2 |\phi|^2 \big) \big( \c^2 g^2 + |q|^2 + |\wt{q}|^2 \big)
\nn \\
%%%
\ & \ \ \ \
- e^2 \Big|
\sqrt{2} \, \wt{q} q
+ \c \, \big( (r^1 + \I r^2) - (s^1 + \I s^2) \big) 
\Big|^2
%\nn \\
%%% 
%\ & \ \ \ \ 
- \frac{e^2}{2} \Big\{ 
\big( |q|^2 - |\wt{q}|^2 \big)
- \sqrt{2} \, \c \big( (- g^2 \wt{r}^3) - t^3 \big) 
\Big\}^2
\nn \\
%%% 
\ & \ \ \ \ 
+ \sqrt{2} \, \c \Big\{ (\chi \wt{\lambda}) - (\ol{\chi} \ol{\wt{\lambda}}) - g^2 (\wt{\chi} \lambda) + g^2 (\ol{\wt{\chi}} \ol{\lambda}{}) \Big\}
%\nn \\
%%%
%\ & \ \ \ \ 
- \I \Big\{ \sigma (\ol{\psi} \psi) - \sigma (\ol{\wt{\psi}} \wt{\psi}) \Big\} 
+ \sqrt{2} \Big\{ \phi (\wt{\psi} \psi) - \ol{\phi} (\ol{\wt{\psi}} \ol{\psi}) \Big\}
\nn \\
%%%
\ & \ \ \ \ 
+ \sqrt{2} \Big\{ 
q (\ol{\psi} \ol{\lambda}) 
+ q (\wt{\psi} \wt{\lambda}) 
- \ol{q} (\psi \lambda) 
- \ol{q} (\ol{\wt{\psi}} \ol{\wt{\lambda}}) 
- \wt{q} (\ol{\wt{\psi}} \ol{\lambda}) 
+ \wt{q} (\psi \wt{\lambda}) 
+ \ol{\wt{q}} (\wt{\psi} \lambda) 
- \ol{\wt{q}}{} (\ol{\psi} \ol{\wt{\lambda}}) 
\Big\}
\, . \label{3d-GLSMKK-comp2}
\end{align}
Note that we introduced a gauge covariant derivative acting on charged fields:
\bsubeq
\begin{alignat}{2}
D_m q
\ &= \ 
\del_m q
+ \I A_m \, q
\, , &\ls
D_m \wt{q}
\ &= \ 
\del_m \wt{q}
- \I A_m \, \wt{q}
\, , \\
%%%%%
D_m \psi_{\alpha}
\ &= \ 
\del_m \psi_{\alpha}
+ \I A_m \, \psi_{\alpha}
\, , &\ls
D_m \wt{\psi}
\ &= \ 
\del_m \wt{\psi}_{\alpha}
- \I A_m \, \wt{\psi}_{\alpha}
\, . 
\end{alignat}
\esubeq
Hereafter, we focus only on the bosonic sector to consider classical vacuum and the low energy theory on it as discussed in the references \cite{Tong:2002rq, Harvey:2005ab, Okuyama:2005gx, Kimura:2013fda}.

%%%%%%%%%%%%%%%%%%%%%%%%%%%%%%%%%
\subsection{Low energy theory} 
\label{sect:SUSY_vacuum}

Let us study supersymmetric low energy theory of the Lagrangian (\ref{3d-GLSMKK-comp2})\footnote{We comment that this discussion is parallel to that in the two-dimensional GLSM for KK5-branes.}.
First, we find three non-trivial equations which 
make vanish the scalar potential:
\bsubeq \label{KK6-SUSY-cond-multi}
\begin{align}
0 \ &= \ 
\big( \sigma^2 + 2 |\phi|^2 \big) \big( \c^2 g^2 + |q|^2 + |\wt{q}|^2 \big)
\, , \label{KK6-pot-03} \\
%%%%%
0 \ &= \ 
\sqrt{2} \, \wt{q} q
+ \c \, \big( (r^1 + \I r^2) - (s^1 + \I s^2) \big) 
\, , \label{KK6-pot-05} \\
%%%%%
0 \ &= \ 
\big( |q|^2 - |\wt{q}|^2 \big)
- \sqrt{2} \, \c \, \big( (- g^2 \wt{r}^3) - t^3 \big) 
\, . \label{KK6-pot-04} 
\end{align}
\esubeq
\smallskip
In this paper we are interested in the Higgs branch where the scalar fields of the hypermultiplets are non-trivial.
Hence, from the equation (\ref{KK6-pot-03}), we can set 
\begin{align}
\sigma \ &= \ 0 \ = \ \phi
\, . \label{KK6-Higgs-multi}
\end{align}
\smallskip
For later convenience, we express $r^3 := -g^2 \wt{r}^3$ for short. 
Analyzing the equations (\ref{KK6-pot-05}) and (\ref{KK6-pot-04}) simultaneously, we can read a non-trivial solution of the charged scalars $(q, \wt{q})$ in such a way that 
\begin{gather}
q
\ = \ 
\frac{\I }{2^{1/4}} 
\sqrt{|\c|} \, \e^{+ \I \alpha} \sqrt{ R + 
\frac{\c}{|\c|}
(r^3 - t^3)}
\, , \ls
%%%%%
\wt{q}
\ = \ 
\frac{\I}{2^{1/4}} 
\frac{\c}{\sqrt{|\c|}}
\, \e^{- \I \alpha} \frac{(r^1 - s^1) + \I (r^2 - s^2)}{\sqrt{ R + \frac{\c}{|\c|} (r^3 - t^3)}}
\, . \label{KK6-qq-multi}
\end{gather}
\smallskip
Here $\alpha$ is a gauge parameter of the Abelian gauge symmetry,
and the value $R$ is defined as 
\begin{align}
2 \big|\sqrt{2} \wt{q} q \big|^2
+ \big( |q|^2 - |\wt{q}|^2 \big)^2
\ &= \ 
2 \c^2 \big\{ (r^1 - s^1)^2 + (r^2 - s^2) + (r^3 - t^3) \big\}
\nn \\
\ &=: \ 
2 \c^2 R^2
\, . 
\end{align}
\smallskip
By using the solution (\ref{KK6-qq-multi}), we can rewrite the kinetic terms of the charged scalar fields $(q, \wt{q})$ in terms of the kinetic terms of the neutral scalar fields $(r^1, r^2, r^3)$ and the gauge fields $A_m$ as follows:
\begin{align}
- |D_m q|^2 
- |D_m \wt{q}|^2
\ &= \ 
- 
\frac{|\c|}{2 \sqrt{2} \, R}
\Big\{ (\del_m r^1)^2 + (\del_m r^2)^2 + (\del_m r^3)^2 \Big\} 
\nn \\
\ & \ \ \ \ 
- 
\sqrt{2} \, |\c| R
\Big\{
A_m
+ \del_m \alpha
+ \frac{1}{\sqrt{2}} \Omega_i \, \del_m r^i
\Big\}^2
\, , \label{KK6-Dqq-multi}
\end{align}
\smallskip
where $\Omega_i$ in the second line is defined as
\begin{gather}
\Omega_i \, \del_m r^i
\ = \ 
\frac{- (r^1 - s^1) \del_m r^2 + (r^2 - s^2) \del_m r^1}{\sqrt{2} \, R (R + \frac{\c}{|\c|} (r^3 - t^3))}
\, . \label{KK6-Omega-multi}
\end{gather}
\vspace{1mm} \\
Substituting the above configuration we obtain 
\begin{align}
\mathcal{L}
\ &= \ 
- \frac{1}{4 e^2} {F}_{mn} {F}^{mn}
- \sqrt{2} \, \c \, {\ve}^{mnp} \, \del_m (B_n A_{p})
\nn \\
%%%
\ & \ \ \ \ 
- \Big( \frac{1}{2 g^2} + 
\frac{|\c|}{2 \sqrt{2} \, R}
\Big) \Big\{ (\del_m r^1)^2 + (\del_m r^2)^2 + (\del_m r^3)^2 \Big\} 
\nn \\
%%%
\ & \ \ \ \ 
- 
\sqrt{2} \, |\c| R
\Big\{
A_m
+ \del_m \alpha
+ \frac{1}{\sqrt{2}} \Omega_i \, \del_m r^i
\Big\}^2
%%%
- \frac{g^2}{2} \Big( \del_m \wt{r}^4 + \sqrt{2} \, \c \, A_m \Big)^2
\nn \\
\ & \ \ \ \ 
+ \text{(fermionic terms)}
\, . \label{3d-GLSMKK-comp3}
\end{align}
\smallskip
Now let us consider the low energy limit.
Since the gauge coupling constant $e$ is dimensionful, 
this grows up to infinity in the IR low energy limit\footnote{Although the FI parameters are also dimensionful, they are protected by the $SU(2)$ R-symmetry and are not changed by virtue of the non-renormalization theorem \cite{Intriligator:1996ex, Kapustin:2010xq}. Hence we can set them to be finite, as in the literature \cite{deBoer:1996mp, Hanany:1996ie, Bullimore:2015lsa}.}.
In this limit the kinetic term of the gauge fields are frozen and they become auxiliary fields.
Then it is easy to solve the equation of motion for the gauge fields:
\begin{align}
A_m
\ &= \ 
- 
\frac{\c}{2 |\c| R H}
\Big( \del_m \wt{\theta} - \c \, \Omega_i \, \del_m r^i \Big)
- \del_m \alpha
- \frac{1}{\sqrt{2}} \Omega_i \, \del_m r^i
\, , \label{sol-A-multi}
\end{align}
\smallskip
where we introduced the following expressions:
\begin{align}
H \ &:= \ 
\frac{1}{g^2} + 
\frac{|\c|}{\sqrt{2} \, R}
\, , \ls
\wt{\theta}
\ := \ 
\wt{r}^4 
- \sqrt{2} \, \c \, \alpha
\, . \label{harm-H}
\end{align}
\smallskip
Substituting this into the Lagrangian (\ref{3d-GLSMKK-comp3}) with $e \to \infty$, we obtain the IR theory as the nonlinear sigma model whose target space is the Taub-NUT space:
\begin{align}
\mathcal{L}_{\text{bos.}}^{\text{IR}}
\ &= \
- \frac{H}{2} \Big\{ (\del_m r^1)^2 + (\del_m r^2)^2 + (\del_m r^3)^2 \Big\} 
- \frac{1}{2 H} \Big( \del_m \wt{\theta} - \c \, \Omega_i \, \del_m r^i \Big)^2
\nn \\
%%%
\ & \ \ \ \ 
- \sqrt{2} \, \c \, {\ve}^{mnp} \, \del_m (B_n \mr{A}_{p})
\, . \label{3d-GLSMKK-IR}
\end{align}
\smallskip
Note that $\mr{A}_{p}$ in the total derivative term is a short expression of the solution (\ref{sol-A-multi}).
The function $H$ is a harmonic function, and $\Omega_{i}$ is the KK-vector.
They are related via $(\nabla \times \Omega)_i = \nabla_i H$ in the three-dimensional space with coordinates $(r^1,r^2,r^3)$.
The variable $\wt{\theta}$ represents nothing but the Taub-NUT circle.
We are able to interpret this model as the ``worldvolume theory of a probe membrane'' moving on the KK6-brane in M-theory.

%%%%%%%%%%%%%%%%%%%%%%%%%%%%%%%%%
\newpage
\subsection{Isometry breaking by instantons}

Now we examine the instantons in the membrane gauge theory.
The bosonic sector of the Lagrangian (\ref{3d-GLSMKK-comp2}) is 
\begin{align}
\mathcal{L}_{\text{bos.}} 
\ &= \
- \frac{1}{e^2} \Big\{ 
\frac{1}{4} {F}_{mn} {F}^{mn}
+ \frac{1}{2} (\del_m \sigma)^2
+ |\del_m \phi|^2
\Big\}
- \sqrt{2} \, \c \, {\ve}^{mnp} \, \del_m (B_n A_{p})
\nn \\
%%%
\ & \ \ \ \ 
- \frac{1}{2 g^2} \Big\{ (\del_m r^1)^2 + (\del_m r^2)^2 + (\del_m (- g^2 \wt{r}^3))^2 \Big\}
- \frac{g^2}{2} \big( \del_m \wt{r}^4 + \sqrt{2} \, \c \, A_m \big)^2
- |D_m q|^2 - |D_m \wt{q}|^2 
\nn \\
%%%
\ & \ \ \ \ 
- \big( \sigma^2 + 2 |\phi|^2 \big) \big( \c^2 g^2 + |q|^2 + |\wt{q}|^2 \big)
- e^2 \Big|
\sqrt{2} \, \wt{q} q
+ \c \, \big( (r^1 - s^1) + \I (r^2 - s^2) \big) 
\Big|^2
\nn \\
%%% 
\ & \ \ \ \ 
- \frac{e^2}{2} \Big\{ \big( |q|^2 - |\wt{q}|^2 \big) - \sqrt{2} \, \c \, (r^3 - t^3) \Big\}^2
\, .  
\end{align}
Following the discussions by \cite{Tong:2002rq, Harvey:2005ab, Kimura:2013zva}, 
we take the limit $g \to 0$ by which the asymptotic radius of the
Taub-NUT circle reduces to zero.
Then the fields $r^1,r^2,r^3$ are frozen and they become constants.
We assume that they are in the vacuum:
\begin{align}
r^1 \ = \ s^1
\, , \quad 
r^2 \ = \ s^2
\, , \quad 
r^3 \ = \ \zeta
\, .
\label{eq:r1r2r3_vev}
\end{align}
Here we take the VEV of $r^3$ as a constant $\zeta$.
By the $SU(2)$ rotation, we make the field $\wt{q}$ be the vacuum configuration $\wt{q} = 0$.
By the same way, we fix $\sigma$, $\phi$ to their VEVs:
\begin{align}
\sigma \ = \ 0 \ = \ \phi
\, . 
\end{align}
In the following, we assume $t^3 = 0$ without loss of generality.
Then we find that the truncate Lagrangian is given by
\begin{align}
\mathcal{L}_{\mathrm{t}} 
\ &= \ 
- \frac{1}{4 e^2} F_{mn} F^{mn} 
- |D_m q|^2
- \frac{e^2}{2} \big( |q|^2 - \sqrt{2} \, \c \, \zeta \big)^2 
- \sqrt{2} \, \c \, {\ve}^{mnp} \del_m (B_n A_p).
\end{align}
This is almost the three-dimensional Abelian-Higgs model but the topological term is 
induced from the BF-like term ${\ve}^{mnp} \del_m (B_n A_p)$
\cite{Brooks:1994nn, Kapustin:1999ha}.

By the Wick rotation $x^0 \to \I x^3$,
we find that the Lagrangian in the Euclidean space is given by 
\begin{align}
\mathcal{L}_{\mathrm{E}} 
\ &= \ 
\frac{1}{4 e^2} F_{mn} F^{mn} + |D_m q|^2 
+ \frac{e^2}{2} \big( |q|^2 - \sqrt{2} \, \c \, \zeta \big)^2
+ \I \sqrt{2} \, \c \, {\varepsilon}^{mnp} \del_m (B_n A_p)
\, ,
\end{align}
where $m,n = 1,2,3$.
Since we are looking for the M2-brane probing the KK6-brane and wrapping along the M-circle, 
one of the three-dimensional worldvolume direction is compactified.
We take the $x^3$-direction as the compactified space and all the fields are 
independent of $x^3$.
We further assume the ansatz:
\begin{align}
A_1, A_2, B_3 \ \neq \ 0
\, , \qquad 
A_3 \ = \ B_1 \ = \ B_2 \ = \ 0
\, , 
\end{align}
and $B_3 = \mathrm{const}$. Then the Euclidean action becomes
\begin{align}
S_{\mathrm{E}} 
\ &= \ 
\int \d^3 x \, 
\Bigg[
\frac{1}{2 e^2} 
\Big\{
F_{12} \pm e^2 
\big( |q|^2 - \sqrt{2} \, \c \, \zeta \big)
\Big\}^2
+ \big| D_1 q \pm \I D_2 q \big|^2 
- \I \sqrt{2} \, \c \, B_3 F_{12} 
\pm \frac{\c \, \zeta}{2 \sqrt{2}} F_{12}
\Bigg].
\end{align}
The Bogomol'nyi bound is saturated if and only if the following equations are satisfied:
\begin{align}
F_{12} 
\pm e^2 \big( |q|^2 - \sqrt{2} \, \c \, \zeta \big) 
\ = \ 0
\, , \qquad 
D_1 q \pm \I D_2 q 
\ = \ 0
\, .
\label{eq:BPSeq}
\end{align}
This is nothing but the Abrikosov-Nielsen-Olesen vortex equations.
The BPS bound is 
\begin{align}
S_{\mathrm{E}} 
\ = \
2 \pi R_{\text{M}} \left( \pm \frac{\c \zeta}{2 \sqrt{2}} - \I \sqrt{2} \, \c \, B_3 \right)
\int \d^2 x \, F_{12}
\, .
\label{eq:KK6_BPS_bound}
\end{align}
Here $2 \pi R_{\text{M}}$ is the length of the M-circle.
From this expression, we find that the isometry along the 
VEV of $B_3$ is broken by the non-zero topological number 
$\textstyle{\frac{1}{2\pi} \int \d^2 x F_{12}} = n$.

Since the three-dimensional Lorentz symmetry is broken on the vortex
\eqref{eq:BPSeq} and the model \eqref{3d-GLSMKK} is reduced to the two-dimensional GLSM for the KK5-brane, 
the instanton calculus is essentially the same with those in
\cite{Harvey:2005ab}. One finds that the instanton induces corrections to the
four fermion coupling which are interpreted as corrections to the
spacetime metric.
The harmonic function in the metric receives the $B_3$ corrections
through the factor $\e^{-S_{\mathrm{E}}}$ with the bound \eqref{eq:KK6_BPS_bound}.
Actually,
this vector field $B_m$ is dual to the scalar field $\wt{r}^4 \sim \wt{\theta}$ in the chiral superfield $\Gamma$.
As we have clarified above, the dimensional reduction unravels that 
the field $B_3$ is interpreted as the
geometric coordinate in the smeared NS5-brane geometry.
From the equation \eqref{eq:KK6_BPS_bound} it is apparent that 
the BPS vortices break the isometry along the $B_3$ direction.
This is the direct analogue of the worldsheet instanton effects \cite{Tong:2002rq}.
Since this $B_3$ appears in the dual side in \eqref{eq:KK6_BPS_bound}, 
it should be interpreted as the dual coordinate corrections to the
KK6-brane.

Finally, we have a short comment. 
In order to investigate instanton corrections in a deeper level, it might be interesting to perform the localization of the three-dimensional gauge theory for KK6-brane (\ref{3d-GLSMKK}),
and compare it with the localization of the two-dimensional GLSM for KK5-brane \cite{Harvey:2014nha, Hiraga:2020lhk} via the double dimensional reduction.

%%%%%%%%%%%%%%%%%%%%%%%%%%%%%%%%%%%%%%%%%%%%%%%%%%%%%%%%%%%%%%%%%%%%%%
%%%%%%%%%%%%%%%%%%%%%%%%%%%%%%%%%%%%%%%%%%%%%%%%%%%%%%%%%%%%%%%%%%%%%%

\section{Conclusion and discussions} \label{sect:conclusion}
In this paper, we studied the KK6-brane localized in the extended space in M-theory.
This analysis is motivated by the T-duality relation between the NS5-brane and the KK5-brane in type IIA string theory.
The winding mode corrections to the KK5-brane is recast in the form of the localization in the winding space.
We examined the KK6-brane solutions in the truncated $E_{7(7)}$
exceptional field theory.
We found that the appropriate coordinate swaps together with the non-trivial scaling
transformations on the M5-brane of codimension four results in the
localized KK6-brane in the extended space.
We found that the localized solution exhibits non-trivial $C$-field components due to the 
dual coordinate dependence.
This is a natural M-theory generalization of the localized KK5-brane in DFT \cite{Kimura:2018hph} where the non-trivial $B$-field is excited due to the winding mode corrections.

We then discussed the physical origin of the dual coordinate dependence of the localized KK6-brane solution.
Although there are no three-cycles in the multi-centered Taub-NUT geometry, the introduction
of the M-circle enables us to define membrane instantons in eleven
dimensions.
We also pointed out that even for the single-centered Taub-NUT space, 
a configuration analogous to the disk instantons for fundamental strings is possible for membranes.
This membrane configuration is expected to induce
the dual coordinate dependence on the KK6-brane geometry.
In order to elucidate the membrane instanton corrections, we next
introduce the three-dimensional gauge theory that describe the probe
M2-brane.

We showed that the three-dimensional $\mathcal{N} = 4$ Abelian gauge
theory is naturally interpreted as a membrane generalization of the
two-dimensional $\mathcal{N} = (4,4)$ GLSM for the fundamental string.
To elucidate this, we performed the dimensional reduction of the
$\mathcal{N} = 4$ theory to two dimensions and find that it gives the
GLSM discussed in \cite{Harvey:2005ab}.
In this process, we found that the vector field in the $\mathcal{N} = 4$ theory plays a key role of the dual coordinate.
In order to find the field configuration that represents the membrane
instantons in the gauge theory, we considered the limit where the asymptotic Taub-NUT circle shrinks to zero \cite{Tong:2002rq, Harvey:2005ab, Kimura:2013zva}.
In this limit, we obtained the truncated gauge model.
We then wrote down the BPS equations for the wrapped membrane and found
that they are vortex equations.
We showed that the topological term is given by the BF-like term and this
induces the violation of the isometry along the dual direction.

We focused on the codimension four solutions in our analysis.
In this case, the instanton calculus is essentially the same with the
ones in the type IIA theory.
The most intrinsic nature of membrane instantons would come from the
localized KK6-brane originating from the non-smeared M5-brane of
codimension five.
Although, we have more components of the $C$-field and the full analysis gets more involved
for the codimension five case, we can extract the metric of the KK6 from EFT.
The result is the one in \eqref{eq:KK6cod3} but the harmonic function
given by
\begin{align}
H = 1 + \frac{h}{(\vec{y}^2 + Y_{wz}^2 + Y_{uz}^2)^{\frac{3}{2}}}.
\end{align}
For such solution, we expect that membrane instanton effects are
captured by a codimension three BPS configuration in the
three-dimensional $\mathcal{N} = 4$ gauge theory. This BPS solution should break the isometry along 
the dual direction $Y_{uz}$ in addition to 
the $Y_{wz}$ direction.
%}

Since EFTs involve many U-duality brane multiplets, there should be the
vast of localized solutions in the extended space.
It would be interesting to study such solutions and discuss physical
origins of wrapping corrections by M-theory branes.
Regarding on the three-dimensional theory, it is interesting to study the U-duality transformation of the membrane gauge theory.
Since the vector field $B_m$ is interpreted as the dual coordinate, we expect that the dual M-theory brane such as the M5-brane geometry
would be described by the field $B_m$. This picture would provide a notable understanding of the dual geometry.
We leave these issues for future works.

%%%%%%%%%%%%%%%%%%%%%%%%%%%%%%%%%%%
\subsection*{Acknowledgments}
The work of T.K. and S.S. is supported in part by Grant-in-Aid for Scientific
Research (C), JSPS KAKENHI Grant Number JP20K03952.
The work of K.S. is supported by Grant-in-Aid for JSPS Research Fellow, JSPS KAKENHI
Grant Number JP20J13957.
%%%%%%%%%%%%%%%%%%%%%%%%%%%%%%%%%%%

\begin{appendix}

\section*{Appendix}

\section{Localized KK6-brane solution}
\label{sect:localized_KK6}

In this appendix, we exhibit the precise form of the localized KK6-brane solution in $E_{7(7)}$ EFT.
After the embedding of \eqref{eq:M5cod4} into \eqref{eq:rescaled_gen_metric} and performing the 
swap of the coordinates \eqref{eq:KK6toM5}, we obtain the explicit form of the generalized metric:
%%%%% modified by TK %%%%%%%%%%%%%%%%%%%
{%\tiny
\scriptsize
\begin{align}
{\d}s_{\mathcal{M}}^2
&= 
\frac{{{\d}u}^2}{H}
+\frac{{{\d}w}^2 C_{{w12}}^2}{H^3}
+\frac{{{\d}w}^2 C_{{w13}}^2}{H^3}
+\frac{{{\d}w}^2 C_{{w23}}^2}{H^3}
+\frac{{{\d}W}_1^2}{H^4}
+\frac{{{\d}W}_2^2}{H^4}
+\frac{{{\d}W}_3^2}{H^4}
+\frac{{{\d}W}_u^2}{H^3}
+\frac{{{\d}W}_v^2}{H^3}
+\frac{{{\d}W}_w^2}{H^3}
+\frac{C_{{zw1}}^2 {{\d}W}_z^2}{H^4}
+\frac{C_{{zw2}}^2 {{\d}W}_z^2}{H^4}
\nn \\
\ & \ \ \
+\frac{C_{{zw3}}^2 {{\d}W}_z^2}{H^4}
+\frac{{{\d}W}_z^2}{H^2}
+\frac{C_{{w12}}^2 {{\d}x}_1^2}{H^2}
+\frac{C_{{w13}}^2 {{\d}x}_1^2}{H^2}
+\frac{C_{{zw1}}^2 {{\d}x}_1^2}{H^2}+{{\d}x}_1^2
+\frac{C_{{w12}}^2 {{\d}x}_2^2}{H^2}
+\frac{C_{{w23}}^2 {{\d}x}_2^2}{H^2}
+\frac{C_{{zw2}}^2 {{\d}x}_2^2}{H^2}+{{\d}x}_2^2
+\frac{C_{{w13}}^2 {{\d}x}_3^2}{H^2}
\nn \\
\ & \ \ \
+\frac{C_{{w23}}^2 {{\d}x}_3^2}{H^2}
+\frac{C_{{zw3}}^2 {{\d}x}_3^2}{H^2}+{{\d}x}_3^2
+\frac{{{\d}Y}_{12}^2}{H^3}
+\frac{{{\d}Y}_{23}^2}{H^3}
+\frac{{{\d}Y}_{31}^2}{H^3}
+\frac{C_{{w23}}^2 {{\d}Y}_{{u1}}^2}{H^4}
+\frac{{{\d}Y}_{{u1}}^2}{H^2}
+\frac{C_{{w13}}^2 {{\d}Y}_{{u2}}^2}{H^4}
+\frac{{{\d}Y}_{{u2}}^2}{H^2}
+\frac{C_{{w12}}^2 {{\d}Y}_{{u3}}^2}{H^4}
\nn \\
\ & \ \ \
+\frac{{{\d}Y}_{{u3}}^2}{H^2}
+\frac{C_{{w12}}^2 {{\d}Y}_{{uv}}^2}{H^3}
+\frac{C_{{w13}}^2 {{\d}Y}_{{uv}}^2}{H^3}
+\frac{C_{{w23}}^2 {{\d}Y}_{{uv}}^2}{H^3}
+\frac{{{\d}Y}_{{uv}}^2}{H}
+\frac{{{\d}Y}_{{uw}}^2}{H}
+\frac{C_{{w12}}^2 {{\d}Y}_{{uz}}^2}{H^2}
+\frac{C_{{w13}}^2 {{\d}Y}_{{uz}}^2}{H^2}
+\frac{C_{{w23}}^2 {{\d}Y}_{{uz}}^2}{H^2}
\nn \\
\ & \ \ \
+\frac{C_{{w23}}^2 C_{{zw1}}^2 {{\d}Y}_{{uz}}^2}{H^4}
+\frac{C_{{zw1}}^2 {{\d}Y}_{{uz}}^2}{H^2}
+\frac{C_{{w13}}^2 C_{{zw2}}^2 {{\d}Y}_{{uz}}^2}{H^4}
+\frac{C_{{zw2}}^2 {{\d}Y}_{{uz}}^2}{H^2}
+\frac{C_{{w12}}^2 C_{{zw3}}^2 {{\d}Y}_{{uz}}^2}{H^4}
+\frac{C_{{zw3}}^2 {{\d}Y}_{{uz}}^2}{H^2}
\nn \\
\ & \ \ \
+\frac{2 C_{{w12}} C_{{w23}} C_{{zw1}} C_{{zw3}} {{\d}Y}_{{uz}}^2}{H^4}
+{{\d}Y}_{{uz}}^2
+\frac{C_{{w23}}^2 {{\d}Y}_{{v1}}^2}{H^4}
+\frac{{{\d}Y}_{{v1}}^2}{H^2}
+\frac{C_{{w13}}^2 {{\d}Y}_{{v2}}^2}{H^4}
+\frac{{{\d}Y}_{{v2}}^2}{H^2}
+\frac{C_{{w12}}^2 {{\d}Y}_{{v3}}^2}{H^4}
+\frac{{{\d}Y}_{{v3}}^2}{H^2}
+\frac{{{\d}Y}_{{vw}}^2}{H}
\nn \\
\ & \ \ \
+\frac{C_{{w12}}^2 {{\d}Y}_{{vz}}^2}{H^2}
+\frac{C_{{w13}}^2 {{\d}Y}_{{vz}}^2}{H^2}
+\frac{C_{{w23}}^2 {{\d}Y}_{{vz}}^2}{H^2}
+\frac{C_{{w23}}^2 C_{{zw1}}^2 {{\d}Y}_{{vz}}^2}{H^4}
+\frac{C_{{zw1}}^2 {{\d}Y}_{{vz}}^2}{H^2}
+\frac{C_{{w13}}^2 C_{{zw2}}^2 {{\d}Y}_{{vz}}^2}{H^4}
+\frac{C_{{zw2}}^2 {{\d}Y}_{{vz}}^2}{H^2}
\nn \\
\ & \ \ \
+\frac{C_{{w12}}^2 C_{{zw3}}^2 {{\d}Y}_{{vz}}^2}{H^4}
+\frac{C_{{zw3}}^2 {{\d}Y}_{{vz}}^2}{H^2}
+\frac{2 C_{{w12}} C_{{w23}} C_{{zw1}} C_{{zw3}} {{\d}Y}_{{vz}}^2}{H^4}
+{{\d}Y}_{{vz}}^2
+\frac{{{\d}Y}_{{w1}}^2}{H^2}
+\frac{{{\d}Y}_{{w2}}^2}{H^2}
+\frac{{{\d}Y}_{{w3}}^2}{H^2}
+\frac{C_{{zw1}}^2 {{\d}Y}_{{wz}}^2}{H^2}
\nn \\
\ & \ \ \
+\frac{C_{{zw2}}^2 {{\d}Y}_{{wz}}^2}{H^2}
+\frac{C_{{zw3}}^2 {{\d}Y}_{{wz}}^2}{H^2}
+{{\d}Y}_{{wz}}^2
+\frac{C_{{w23}}^2 {{\d}Y}_{{z1}}^2}{H^3}
+\frac{C_{{zw2}}^2 {{\d}Y}_{{z1}}^2}{H^3}
+\frac{C_{{zw3}}^2 {{\d}Y}_{{z1}}^2}{H^3}
+\frac{{{\d}Y}_{{z1}}^2}{H}
+\frac{C_{{w13}}^2 {{\d}Y}_{{z2}}^2}{H^3}
+\frac{C_{{zw1}}^2 {{\d}Y}_{{z2}}^2}{H^3}
\nn \\
\ & \ \ \
+\frac{C_{{zw3}}^2 {{\d}Y}_{{z2}}^2}{H^3}
+\frac{{{\d}Y}_{{z2}}^2}{H}
+\frac{C_{{w12}}^2 {{\d}Y}_{{z3}}^2}{H^3}
+\frac{C_{{zw1}}^2 {{\d}Y}_{{z3}}^2}{H^3}
+\frac{C_{{zw2}}^2 {{\d}Y}_{{z3}}^2}{H^3}
+\frac{{{\d}Y}_{{z3}}^2}{H}
+\frac{C_{{w12}}^2 {{\d}Z}_{12}^2}{H^3}
+\frac{C_{{zw1}}^2 {{\d}Z}_{12}^2}{H^3}
+\frac{C_{{zw2}}^2 {{\d}Z}_{12}^2}{H^3}
\nn \\
\ & \ \ \
+\frac{{{\d}Z}_{12}^2}{H}
+\frac{C_{{w23}}^2 {{\d}Z}_{23}^2}{H^3}
+\frac{C_{{zw2}}^2 {{\d}Z}_{23}^2}{H^3}
+\frac{C_{{zw3}}^2 {{\d}Z}_{23}^2}{H^3}
+\frac{{{\d}Z}_{23}^2}{H}
+\frac{C_{{w13}}^2 {{\d}Z}_{31}^2}{H^3}
+\frac{C_{{zw1}}^2 {{\d}Z}_{31}^2}{H^3}
+\frac{C_{{zw3}}^2 {{\d}Z}_{31}^2}{H^3}
+\frac{{{\d}Z}_{31}^2}{H}
+\frac{C_{{zw1}}^2 {{\d}Z}_{{u1}}^2}{H^4}
\nn \\
\ & \ \ \
+\frac{{{\d}Z}_{{u1}}^2}{H^2}
+\frac{C_{{zw2}}^2 {{\d}Z}_{{u2}}^2}{H^4}
+\frac{{{\d}Z}_{{u2}}^2}{H^2}
+\frac{C_{{zw3}}^2 {{\d}Z}_{{u3}}^2}{H^4}
+\frac{{{\d}Z}_{{u3}}^2}{H^2}
+\frac{{{\d}Z}_{{uv}}^2}{H^3}
+\frac{{{\d}Z}_{{uw}}^2}{H^3}
+\frac{{{\d}Z}_{{uz}}^2}{H^4}
+\frac{C_{{zw1}}^2 {{\d}Z}_{{v1}}^2}{H^4}
+\frac{{{\d}Z}_{{v1}}^2}{H^2}
+\frac{C_{{zw2}}^2 {{\d}Z}_{{v2}}^2}{H^4}
\nn \\
\ & \ \ \
+\frac{{{\d}Z}_{{v2}}^2}{H^2}
+\frac{C_{{zw3}}^2 {{\d}Z}_{{v3}}^2}{H^4}
+\frac{{{\d}Z}_{{v3}}^2}{H^2}
+\frac{{{\d}Z}_{{vw}}^2}{H^3}
+\frac{{{\d}Z}_{{vz}}^2}{H^4}
+\frac{C_{{w12}}^2 {{\d}Z}_{{w1}}^2}{H^4}
+\frac{C_{{w13}}^2 {{\d}Z}_{{w1}}^2}{H^4}
+\frac{C_{{zw1}}^2 {{\d}Z}_{{w1}}^2}{H^4}
+\frac{{{\d}Z}_{{w1}}^2}{H^2}
+\frac{C_{{w12}}^2 {{\d}Z}_{{w2}}^2}{H^4}
\nn \\
\ & \ \ \
+\frac{C_{{w23}}^2 {{\d}Z}_{{w2}}^2}{H^4}
+\frac{C_{{zw2}}^2 {{\d}Z}_{{w2}}^2}{H^4}
+\frac{{{\d}Z}_{{w2}}^2}{H^2}
+\frac{C_{{w13}}^2 {{\d}Z}_{{w3}}^2}{H^4}
+\frac{C_{{w23}}^2 {{\d}Z}_{{w3}}^2}{H^4}
+\frac{C_{{zw3}}^2 {{\d}Z}_{{w3}}^2}{H^4}
+\frac{{{\d}Z}_{{w3}}^2}{H^2}
+\frac{{{\d}Z}_{{wz}}^2}{H^4}
+\frac{{{\d}Z}_{{z1}}^2}{H^3}
+\frac{{{\d}Z}_{{z2}}^2}{H^3}
\nn \\
\ & \ \ \
+\frac{{{\d}Z}_{{z3}}^2}{H^3}
+\frac{2 {{\d}z} C_{{zw1}} {{\d}x}_1}{H^2}
+\frac{2 {{\d}z} C_{{zw2}} {{\d}x}_2}{H^2}
+\frac{2 C_{{w13}} C_{{w23}} {{\d}x}_1 {{\d}x}_2}{H^2}
+\frac{2 C_{{zw1}} C_{{zw2}} {{\d}x}_1 {{\d}x}_2}{H^2}
+\frac{2 {{\d}z} C_{{zw3}} {{\d}x}_3}{H^2}
\nn \\
\ & \ \ \
+\frac{2 C_{{zw1}} C_{{zw3}} {{\d}x}_1 {{\d}x}_3}{H^2}
+\frac{2 C_{{w12}} C_{{w13}} {{\d}x}_2 {{\d}x}_3}{H^2}
+\frac{2 C_{{zw2}} C_{{zw3}} {{\d}x}_2 {{\d}x}_3}{H^2}
+\frac{2 {{\d}w} C_{{w12}} {{\d}Y}_{12}}{H^3}
+\frac{2 {{\d}w} C_{{w23}} {{\d}Y}_{23}}{H^3}
\nn \\
\ & \ \ \
+\frac{2 C_{{w12}} C_{{w23}} {{\d}Y}_{{u1}} {{\d}Y}_{{u3}}}{H^4}
+\frac{2 C_{{w13}} C_{{w23}} C_{{zw2}} {{\d}Y}_{{u1}} {{\d}Y}_{{uz}}}{H^4}
+\frac{2 C_{{w13}} C_{{w23}} C_{{zw1}} {{\d}Y}_{{u2}} {{\d}Y}_{{uz}}}{H^4}
+\frac{2 C_{{w12}} C_{{w13}} C_{{zw3}} {{\d}Y}_{{u2}} {{\d}Y}_{{uz}}}{H^4}
\nn \\
\ & \ \ \
+\frac{2 C_{{w12}} C_{{w13}} C_{{zw2}} {{\d}Y}_{{u3}} {{\d}Y}_{{uz}}}{H^4}
+\frac{2 C_{{w12}} C_{{w23}} {{\d}Y}_{{v1}} {{\d}Y}_{{v3}}}{H^4}
+\frac{2 C_{{w13}} C_{{w23}} C_{{zw2}} {{\d}Y}_{{v1}} {{\d}Y}_{{vz}}}{H^4}
+\frac{2 C_{{w13}} C_{{w23}} C_{{zw1}} {{\d}Y}_{{v2}} {{\d}Y}_{{vz}}}{H^4}
\nn \\
\ & \ \ \
+\frac{2 C_{{w12}} C_{{w13}} C_{{zw3}} {{\d}Y}_{{v2}} {{\d}Y}_{{vz}}}{H^4}
+\frac{2 C_{{w12}} C_{{w13}} C_{{zw2}} {{\d}Y}_{{v3}} {{\d}Y}_{{vz}}}{H^4}
+\frac{2 C_{{w12}} {{\d}x}_2 {{\d}Y}_{{w1}}}{H^2}
+\frac{2 C_{{w13}} {{\d}x}_3 {{\d}Y}_{{w1}}}{H^2}
+\frac{2 C_{{w23}} {{\d}x}_3 {{\d}Y}_{{w2}}}{H^2}
\nn \\
\ & \ \ \
+\frac{2 C_{{w12}} C_{{zw2}} {{\d}x}_1 {{\d}Y}_{{wz}}}{H^2}
+\frac{2 C_{{w13}} C_{{zw3}} {{\d}x}_1 {{\d}Y}_{{wz}}}{H^2}
+\frac{2 C_{{w23}} C_{{zw3}} {{\d}x}_2 {{\d}Y}_{{wz}}}{H^2}
+\frac{2 {{\d}w} C_{{w12}} C_{{zw2}} {{\d}Y}_{{z1}}}{H^3}
+\frac{2 {{\d}w} C_{{w13}} C_{{zw3}} {{\d}Y}_{{z1}}}{H^3}
\nn \\
\ & \ \ \
+\frac{2 C_{{zw2}} {{\d}Y}_{12} {{\d}Y}_{{z1}}}{H^3}
+\frac{2 {{\d}w} C_{{w23}} C_{{zw3}} {{\d}Y}_{{z2}}}{H^3}
+\frac{2 C_{{zw3}} {{\d}Y}_{23} {{\d}Y}_{{z2}}}{H^3}
+\frac{2 C_{{zw1}} {{\d}Y}_{31} {{\d}Y}_{{z3}}}{H^3}
+\frac{2 C_{{w12}} C_{{w23}} {{\d}Y}_{{z1}} {{\d}Y}_{{z3}}}{H^3}
\nn \\
\ & \ \ \
+\frac{2 C_{{w12}} {{\d}W}_w {{\d}Z}_{12}}{H^3}
+\frac{2 C_{{w13}} C_{{zw1}} {{\d}Y}_{{uv}} {{\d}Z}_{12}}{H^3}
+\frac{2 C_{{w23}} C_{{zw2}} {{\d}Y}_{{uv}} {{\d}Z}_{12}}{H^3}
+\frac{2 C_{{w23}} {{\d}W}_w {{\d}Z}_{23}}{H^3}
+\frac{2 C_{{w12}} C_{{w23}} {{\d}Z}_{12} {{\d}Z}_{23}}{H^3}
\nn \\
\ & \ \ \
+\frac{2 C_{{w12}} C_{{zw1}} {{\d}Y}_{{uv}} {{\d}Z}_{31}}{H^3}
+\frac{2 C_{{w23}} C_{{zw1}} {{\d}Y}_{{v1}} {{\d}Z}_{{u1}}}{H^4}
+\frac{2 C_{{w12}} C_{{zw1}} {{\d}Y}_{{v3}} {{\d}Z}_{{u1}}}{H^4}
+\frac{2 C_{{w13}} C_{{zw1}} C_{{zw2}} {{\d}Y}_{{vz}} {{\d}Z}_{{u1}}}{H^4}
\nn \\
\ & \ \ \
+\frac{2 C_{{w23}} C_{{zw2}} {{\d}Y}_{{v1}} {{\d}Z}_{{u2}}}{H^4}
+\frac{2 C_{{w12}} C_{{zw2}} {{\d}Y}_{{v3}} {{\d}Z}_{{u2}}}{H^4}
+\frac{2 C_{{w13}} C_{{zw2}}^2 {{\d}Y}_{{vz}} {{\d}Z}_{{u2}}}{H^4}
+\frac{2 C_{{w13}} {{\d}Y}_{{vz}} {{\d}Z}_{{u2}}}{H^2}
+\frac{2 C_{{zw1}} C_{{zw2}} {{\d}Z}_{{u1}} {{\d}Z}_{{u2}}}{H^4}
\nn \\
\ & \ \ \
+\frac{2 C_{{w23}} C_{{zw3}} {{\d}Y}_{{v1}} {{\d}Z}_{{u3}}}{H^4}
+\frac{2 C_{{w12}} C_{{zw3}} {{\d}Y}_{{v3}} {{\d}Z}_{{u3}}}{H^4}
+\frac{2 C_{{w13}} C_{{zw2}} C_{{zw3}} {{\d}Y}_{{vz}} {{\d}Z}_{{u3}}}{H^4}
+\frac{2 C_{{zw1}} C_{{zw3}} {{\d}Z}_{{u1}} {{\d}Z}_{{u3}}}{H^4}
\nn \\
\ & \ \ \
+\frac{2 C_{{zw2}} C_{{zw3}} {{\d}Z}_{{u2}} {{\d}Z}_{{u3}}}{H^4}
+\frac{2 C_{{w13}} {{\d}Y}_{{z2}} {{\d}Z}_{{uv}}}{H^3}
+\frac{2 C_{{w23}} {{\d}Y}_{{v1}} {{\d}Z}_{{uz}}}{H^4}
+\frac{2 C_{{w12}} {{\d}Y}_{{v3}} {{\d}Z}_{{uz}}}{H^4}
+\frac{2 C_{{w13}} C_{{zw2}} {{\d}Y}_{{vz}} {{\d}Z}_{{uz}}}{H^4}
\nn \\
\ & \ \ \
+\frac{2 C_{{zw1}} {{\d}Z}_{{u1}} {{\d}Z}_{{uz}}}{H^4}
+\frac{2 C_{{zw2}} {{\d}Z}_{{u2}} {{\d}Z}_{{uz}}}{H^4}
+\frac{2 C_{{zw3}} {{\d}Z}_{{u3}} {{\d}Z}_{{uz}}}{H^4}
+\frac{2 C_{{w13}} C_{{zw1}} {{\d}Y}_{{u2}} {{\d}Z}_{{v1}}}{H^4}
+\frac{2 C_{{w23}} C_{{zw1}}^2 {{\d}Y}_{{uz}} {{\d}Z}_{{v1}}}{H^4}
\nn \\
\ & \ \ \
+\frac{2 C_{{w23}} {{\d}Y}_{{uz}} {{\d}Z}_{{v1}}}{H^2}
+\frac{2 C_{{w12}} C_{{zw1}} C_{{zw3}} {{\d}Y}_{{uz}} {{\d}Z}_{{v1}}}{H^4}
+\frac{2 C_{{w13}} C_{{zw2}} {{\d}Y}_{{u2}} {{\d}Z}_{{v2}}}{H^4}
+\frac{2 C_{{w23}} C_{{zw1}} C_{{zw2}} {{\d}Y}_{{uz}} {{\d}Z}_{{v2}}}{H^4}
\nn \\
\ & \ \ \
+\frac{2 C_{{w12}} C_{{zw2}} C_{{zw3}} {{\d}Y}_{{uz}} {{\d}Z}_{{v2}}}{H^4}
+\frac{2 C_{{zw1}} C_{{zw2}} {{\d}Z}_{{v1}} {{\d}Z}_{{v2}}}{H^4}
+\frac{2 C_{{w13}} C_{{zw3}} {{\d}Y}_{{u2}} {{\d}Z}_{{v3}}}{H^4}
+\frac{2 C_{{w12}} C_{{zw3}}^2 {{\d}Y}_{{uz}} {{\d}Z}_{{v3}}}{H^4}
\nn \\
\ & \ \ \
+\frac{2 C_{{w12}} {{\d}Y}_{{uz}} {{\d}Z}_{{v3}}}{H^2}
+\frac{2 C_{{w23}} C_{{zw1}} C_{{zw3}} {{\d}Y}_{{uz}} {{\d}Z}_{{v3}}}{H^4}
+\frac{2 C_{{zw1}} C_{{zw3}} {{\d}Z}_{{v1}} {{\d}Z}_{{v3}}}{H^4}
+\frac{2 C_{{zw2}} C_{{zw3}} {{\d}Z}_{{v2}} {{\d}Z}_{{v3}}}{H^4}
\nn \\
\ & \ \ \
+\frac{2 C_{{w13}} {{\d}Y}_{{u2}} {{\d}Z}_{{vz}}}{H^4}
+\frac{2 C_{{w23}} C_{{zw1}} {{\d}Y}_{{uz}} {{\d}Z}_{{vz}}}{H^4}
+\frac{2 C_{{w12}} C_{{zw3}} {{\d}Y}_{{uz}} {{\d}Z}_{{vz}}}{H^4}
+\frac{2 C_{{zw1}} {{\d}Z}_{{v1}} {{\d}Z}_{{vz}}}{H^4}
+\frac{2 C_{{zw2}} {{\d}Z}_{{v2}} {{\d}Z}_{{vz}}}{H^4}
\nn \\
\ & \ \ \
+\frac{2 C_{{zw3}} {{\d}Z}_{{v3}} {{\d}Z}_{{vz}}}{H^4}
+\frac{2 C_{{w12}} {{\d}W}_2 {{\d}Z}_{{w1}}}{H^4}
+\frac{2 C_{{w13}} {{\d}W}_3 {{\d}Z}_{{w1}}}{H^4}
+\frac{2 C_{{w23}} {{\d}W}_3 {{\d}Z}_{{w2}}}{H^4}
+\frac{2 C_{{w12}} C_{{zw1}} {{\d}W}_z {{\d}Z}_{{w2}}}{H^4}
\nn \\
\ & \ \ \
+\frac{2 C_{{w13}} C_{{w23}} {{\d}Z}_{{w1}} {{\d}Z}_{{w2}}}{H^4}
+\frac{2 C_{{zw1}} C_{{zw2}} {{\d}Z}_{{w1}} {{\d}Z}_{{w2}}}{H^4}
+\frac{2 C_{{w13}} C_{{zw1}} {{\d}W}_z {{\d}Z}_{{w3}}}{H^4}
+\frac{2 C_{{w23}} C_{{zw2}} {{\d}W}_z {{\d}Z}_{{w3}}}{H^4}
\nn \\
\ & \ \ \
+\frac{2 C_{{zw1}} C_{{zw3}} {{\d}Z}_{{w1}} {{\d}Z}_{{w3}}}{H^4}
+\frac{2 C_{{w12}} C_{{w13}} {{\d}Z}_{{w2}} {{\d}Z}_{{w3}}}{H^4}
+\frac{2 C_{{zw2}} C_{{zw3}} {{\d}Z}_{{w2}} {{\d}Z}_{{w3}}}{H^4}
+\frac{2 C_{{zw1}} {{\d}Z}_{{w1}} {{\d}Z}_{{wz}}}{H^4}
+\frac{2 C_{{zw2}} {{\d}Z}_{{w2}} {{\d}Z}_{{wz}}}{H^4}
\nn \\
\ & \ \ \
+\frac{2 C_{{zw3}} {{\d}Z}_{{w3}} {{\d}Z}_{{wz}}}{H^4}
+\frac{2 C_{{zw3}} {{\d}Z}_{31} {{\d}Z}_{{z1}}}{H^3}
+\frac{2 C_{{w13}} {{\d}Y}_{{uv}} {{\d}Z}_{{z2}}}{H^3}
+\frac{2 C_{{zw1}} {{\d}Z}_{12} {{\d}Z}_{{z2}}}{H^3}
+\frac{2 C_{{zw2}} {{\d}Z}_{23} {{\d}Z}_{{z3}}}{H^3}
+\frac{{{\d}v}^2}{H}
+\frac{{{\d}w}^2}{H}
+\frac{{{\d}z}^2}{H^2}
\nn \\
\ & \ \ \
-\frac{2 C_{{w12}} C_{{w23}} {{\d}x}_1 {{\d}x}_3}{H^2}
-\frac{2 C_{{zw1}} {{\d}Y}_{{u1}} {{\d}Y}_{{uz}}}{H^2}
-\frac{2 C_{{zw2}} {{\d}Y}_{{u2}} {{\d}Y}_{{uz}}}{H^2}
-\frac{2 C_{{zw3}} {{\d}Y}_{{u3}} {{\d}Y}_{{uz}}}{H^2}
-\frac{2 C_{{zw1}} {{\d}Y}_{{v1}} {{\d}Y}_{{vz}}}{H^2}
\nn \\
\ & \ \ \
-\frac{2 C_{{zw2}} {{\d}Y}_{{v2}} {{\d}Y}_{{vz}}}{H^2}
-\frac{2 C_{{zw3}} {{\d}Y}_{{v3}} {{\d}Y}_{{vz}}}{H^2}
-\frac{2 C_{{w12}} {{\d}x}_1 {{\d}Y}_{{w2}}}{H^2}
-\frac{2 C_{{w13}} {{\d}x}_1 {{\d}Y}_{{w3}}}{H^2}
-\frac{2 C_{{w23}} {{\d}x}_2 {{\d}Y}_{{w3}}}{H^2}
-\frac{2 C_{{w12}} C_{{zw1}} {{\d}x}_2 {{\d}Y}_{{wz}}}{H^2}
\nn \\
\ & \ \ \
-\frac{2 C_{{w13}} C_{{zw1}} {{\d}x}_3 {{\d}Y}_{{wz}}}{H^2}
-\frac{2 C_{{w23}} C_{{zw2}} {{\d}x}_3 {{\d}Y}_{{wz}}}{H^2}
-\frac{2 C_{{zw1}} {{\d}Y}_{{w1}} {{\d}Y}_{{wz}}}{H^2}
-\frac{2 C_{{zw2}} {{\d}Y}_{{w2}} {{\d}Y}_{{wz}}}{H^2}
-\frac{2 C_{{zw3}} {{\d}Y}_{{w3}} {{\d}Y}_{{wz}}}{H^2}
\nn \\
\ & \ \ \
-\frac{2 C_{{w23}} {{\d}Y}_{{vz}} {{\d}Z}_{{u1}}}{H^2}
-\frac{2 C_{{w12}} {{\d}Y}_{{vz}} {{\d}Z}_{{u3}}}{H^2}
-\frac{2 C_{{w13}} {{\d}Y}_{{uz}} {{\d}Z}_{{v2}}}{H^2}
-\frac{2 {{\d}w} C_{{w13}} {{\d}Y}_{31}}{H^3}
-\frac{2 C_{{zw3}} {{\d}Y}_{31} {{\d}Y}_{{z1}}}{H^3}
-\frac{2 {{\d}w} C_{{w12}} C_{{zw1}} {{\d}Y}_{{z2}}}{H^3}
\nn \\
\ & \ \ \
-\frac{2 C_{{zw1}} {{\d}Y}_{12} {{\d}Y}_{{z2}}}{H^3}
-\frac{2 C_{{w13}} C_{{w23}} {{\d}Y}_{{z1}} {{\d}Y}_{{z2}}}{H^3}
-\frac{2 C_{{zw1}} C_{{zw2}} {{\d}Y}_{{z1}} {{\d}Y}_{{z2}}}{H^3}
-\frac{2 {{\d}w} C_{{w13}} C_{{zw1}} {{\d}Y}_{{z3}}}{H^3}
-\frac{2 {{\d}w} C_{{w23}} C_{{zw2}} {{\d}Y}_{{z3}}}{H^3}
\nn \\
\ & \ \ \
-\frac{2 C_{{zw2}} {{\d}Y}_{23} {{\d}Y}_{{z3}}}{H^3}
-\frac{2 C_{{zw1}} C_{{zw3}} {{\d}Y}_{{z1}} {{\d}Y}_{{z3}}}{H^3}
-\frac{2 C_{{w12}} C_{{w13}} {{\d}Y}_{{z2}} {{\d}Y}_{{z3}}}{H^3}
-\frac{2 C_{{zw2}} C_{{zw3}} {{\d}Y}_{{z2}} {{\d}Y}_{{z3}}}{H^3}
-\frac{2 C_{{w12}} C_{{zw2}} {{\d}Y}_{{uv}} {{\d}Z}_{23}}{H^3}
\nn \\
\ & \ \ \
-\frac{2 C_{{w13}} C_{{zw3}} {{\d}Y}_{{uv}} {{\d}Z}_{23}}{H^3}
-\frac{2 C_{{zw1}} C_{{zw3}} {{\d}Z}_{12} {{\d}Z}_{23}}{H^3}
-\frac{2 C_{{w13}} {{\d}W}_w {{\d}Z}_{31}}{H^3}
-\frac{2 C_{{w23}} C_{{zw3}} {{\d}Y}_{{uv}} {{\d}Z}_{31}}{H^3}
-\frac{2 C_{{w12}} C_{{w13}} {{\d}Z}_{12} {{\d}Z}_{31}}{H^3}
\nn \\
\ & \ \ \
-\frac{2 C_{{zw2}} C_{{zw3}} {{\d}Z}_{12} {{\d}Z}_{31}}{H^3}
-\frac{2 C_{{w13}} C_{{w23}} {{\d}Z}_{23} {{\d}Z}_{31}}{H^3}
-\frac{2 C_{{zw1}} C_{{zw2}} {{\d}Z}_{23} {{\d}Z}_{31}}{H^3}
-\frac{2 C_{{w23}} {{\d}Y}_{{z1}} {{\d}Z}_{{uv}}}{H^3}
-\frac{2 C_{{w12}} {{\d}Y}_{{z3}} {{\d}Z}_{{uv}}}{H^3}
\nn \\
\ & \ \ \
-\frac{2 C_{{w23}} {{\d}Y}_{{uv}} {{\d}Z}_{{z1}}}{H^3}
-\frac{2 C_{{zw2}} {{\d}Z}_{12} {{\d}Z}_{{z1}}}{H^3}
-\frac{2 C_{{zw3}} {{\d}Z}_{23} {{\d}Z}_{{z2}}}{H^3}
-\frac{2 C_{{w12}} {{\d}Y}_{{uv}} {{\d}Z}_{{z3}}}{H^3}
-\frac{2 C_{{zw1}} {{\d}Z}_{31} {{\d}Z}_{{z3}}}{H^3}
\nn \\
\ & \ \ \
-\frac{2 C_{{w13}} C_{{w23}} C_{{zw1}} C_{{zw2}} {{\d}Y}_{{uz}}^2}{H^4}
-\frac{2 C_{{w12}} C_{{w13}} C_{{zw2}} C_{{zw3}} {{\d}Y}_{{uz}}^2}{H^4}
-\frac{2 C_{{w13}} C_{{w23}} C_{{zw1}} C_{{zw2}} {{\d}Y}_{{vz}}^2}{H^4}
-\frac{2 C_{{w12}} C_{{w13}} C_{{zw2}} C_{{zw3}} {{\d}Y}_{{vz}}^2}{H^4}
\nn \\
\ & \ \ \
-\frac{2 C_{{zw1}} {{\d}W}_1 {{\d}W}_z}{H^4}
-\frac{2 C_{{zw2}} {{\d}W}_2 {{\d}W}_z}{H^4}
-\frac{2 C_{{zw3}} {{\d}W}_3 {{\d}W}_z}{H^4}
-\frac{2 C_{{w13}} C_{{w23}} {{\d}Y}_{{u1}} {{\d}Y}_{{u2}}}{H^4}
-\frac{2 C_{{w12}} C_{{w13}} {{\d}Y}_{{u2}} {{\d}Y}_{{u3}}}{H^4}
\nn \\
\ & \ \ \
-\frac{2 C_{{w23}}^2 C_{{zw1}} {{\d}Y}_{{u1}} {{\d}Y}_{{uz}}}{H^4}
-\frac{2 C_{{w12}} C_{{w23}} C_{{zw3}} {{\d}Y}_{{u1}} {{\d}Y}_{{uz}}}{H^4}
-\frac{2 C_{{w13}}^2 C_{{zw2}} {{\d}Y}_{{u2}} {{\d}Y}_{{uz}}}{H^4}
-\frac{2 C_{{w12}} C_{{w23}} C_{{zw1}} {{\d}Y}_{{u3}} {{\d}Y}_{{uz}}}{H^4}
\nn \\
\ & \ \ \
-\frac{2 C_{{w12}}^2 C_{{zw3}} {{\d}Y}_{{u3}} {{\d}Y}_{{uz}}}{H^4}
-\frac{2 C_{{w13}} C_{{w23}} {{\d}Y}_{{v1}} {{\d}Y}_{{v2}}}{H^4}
-\frac{2 C_{{w12}} C_{{w13}} {{\d}Y}_{{v2}} {{\d}Y}_{{v3}}}{H^4}
-\frac{2 C_{{w23}}^2 C_{{zw1}} {{\d}Y}_{{v1}} {{\d}Y}_{{vz}}}{H^4}
\nn \\
\ & \ \ \ 
-\frac{2 C_{{w12}} C_{{w23}} C_{{zw3}} {{\d}Y}_{{v1}} {{\d}Y}_{{vz}}}{H^4}
-\frac{2 C_{{w13}}^2 C_{{zw2}} {{\d}Y}_{{v2}} {{\d}Y}_{{vz}}}{H^4}
-\frac{2 C_{{w12}} C_{{w23}} C_{{zw1}} {{\d}Y}_{{v3}} {{\d}Y}_{{vz}}}{H^4}
-\frac{2 C_{{w12}}^2 C_{{zw3}} {{\d}Y}_{{v3}} {{\d}Y}_{{vz}}}{H^4}
\nn \\
\ & \ \ \
-\frac{2 C_{{w13}} C_{{zw1}} {{\d}Y}_{{v2}} {{\d}Z}_{{u1}}}{H^4}
-\frac{2 C_{{w23}} C_{{zw1}}^2 {{\d}Y}_{{vz}} {{\d}Z}_{{u1}}}{H^4}
-\frac{2 C_{{w12}} C_{{zw1}} C_{{zw3}} {{\d}Y}_{{vz}} {{\d}Z}_{{u1}}}{H^4}
-\frac{2 C_{{w13}} C_{{zw2}} {{\d}Y}_{{v2}} {{\d}Z}_{{u2}}}{H^4}
\nn \\
\ & \ \ \
-\frac{2 C_{{w23}} C_{{zw1}} C_{{zw2}} {{\d}Y}_{{vz}} {{\d}Z}_{{u2}}}{H^4}
-\frac{2 C_{{w12}} C_{{zw2}} C_{{zw3}} {{\d}Y}_{{vz}} {{\d}Z}_{{u2}}}{H^4}
-\frac{2 C_{{w13}} C_{{zw3}} {{\d}Y}_{{v2}} {{\d}Z}_{{u3}}}{H^4}
-\frac{2 C_{{w12}} C_{{zw3}}^2 {{\d}Y}_{{vz}} {{\d}Z}_{{u3}}}{H^4}
\nn \\
\ & \ \ \
-\frac{2 C_{{w23}} C_{{zw1}} C_{{zw3}} {{\d}Y}_{{vz}} {{\d}Z}_{{u3}}}{H^4}
-\frac{2 C_{{w13}} {{\d}Y}_{{v2}} {{\d}Z}_{{uz}}}{H^4}
-\frac{2 C_{{w23}} C_{{zw1}} {{\d}Y}_{{vz}} {{\d}Z}_{{uz}}}{H^4}
-\frac{2 C_{{w12}} C_{{zw3}} {{\d}Y}_{{vz}} {{\d}Z}_{{uz}}}{H^4}
\nn \\
\ & \ \ \
-\frac{2 C_{{w23}} C_{{zw1}} {{\d}Y}_{{u1}} {{\d}Z}_{{v1}}}{H^4}
-\frac{2 C_{{w12}} C_{{zw1}} {{\d}Y}_{{u3}} {{\d}Z}_{{v1}}}{H^4}
-\frac{2 C_{{w13}} C_{{zw1}} C_{{zw2}} {{\d}Y}_{{uz}} {{\d}Z}_{{v1}}}{H^4}
-\frac{2 C_{{w23}} C_{{zw2}} {{\d}Y}_{{u1}} {{\d}Z}_{{v2}}}{H^4}
\nn \\
\ & \ \ \
-\frac{2 C_{{w12}} C_{{zw2}} {{\d}Y}_{{u3}} {{\d}Z}_{{v2}}}{H^4}
-\frac{2 C_{{w13}} C_{{zw2}}^2 {{\d}Y}_{{uz}} {{\d}Z}_{{v2}}}{H^4}
-\frac{2 C_{{w23}} C_{{zw3}} {{\d}Y}_{{u1}} {{\d}Z}_{{v3}}}{H^4}
-\frac{2 C_{{w12}} C_{{zw3}} {{\d}Y}_{{u3}} {{\d}Z}_{{v3}}}{H^4}
\nn \\
\ & \ \ \
-\frac{2 C_{{w13}} C_{{zw2}} C_{{zw3}} {{\d}Y}_{{uz}} {{\d}Z}_{{v3}}}{H^4}
-\frac{2 C_{{w23}} {{\d}Y}_{{u1}} {{\d}Z}_{{vz}}}{H^4}
-\frac{2 C_{{w12}} {{\d}Y}_{{u3}} {{\d}Z}_{{vz}}}{H^4}
-\frac{2 C_{{w13}} C_{{zw2}} {{\d}Y}_{{uz}} {{\d}Z}_{{vz}}}{H^4}
-\frac{2 C_{{w12}} C_{{zw2}} {{\d}W}_z {{\d}Z}_{{w1}}}{H^4}
\nn \\
\ & \ \ \
-\frac{2 C_{{w13}} C_{{zw3}} {{\d}W}_z {{\d}Z}_{{w1}}}{H^4}
-\frac{2 C_{{w12}} {{\d}W}_1 {{\d}Z}_{{w2}}}{H^4}
-\frac{2 C_{{w23}} C_{{zw3}} {{\d}W}_z {{\d}Z}_{{w2}}}{H^4}
-\frac{2 C_{{w13}} {{\d}W}_1 {{\d}Z}_{{w3}}}{H^4}
-\frac{2 C_{{w23}} {{\d}W}_2 {{\d}Z}_{{w3}}}{H^4}
\nn \\
\ & \ \ \
-\frac{2 C_{{w12}} C_{{w23}} {{\d}Z}_{{w1}} {{\d}Z}_{{w3}}}{H^4}.
\refstepcounter{equation}{\label{eq:gm_mathematica}}
\tag*{\normalsize (\theequation)}
%\label{eq:gm_mathematica}
\end{align}
}%
%%%%% modified by TK, end %%%%%%%%%%%%%%%%%%%
We work in the Mathematica to calculate this result.
From the $g^{-1} g^{\mu \nu} \d W_{\mu} \d W_{\nu}$ sector, we determine the 
scale factor $\e^{2 \alpha'_3} = 1$ and the metric as 
\begin{align}
\d s^2 
\ &= \ 
\d u^2 + \d v^2 + \d w^2 + H \d \vec{y}^{\, 2} + H^{-1} (\d z + C_{izw} \, \d y^i)^2
\, .
\label{eq:lKK6_metric}
\end{align}
From the $g^{-\frac{1}{2}} C_{\mu \nu \alpha} g^{\alpha \beta} \d Z^{\mu \nu} \d W_{\rho}$ sector, 
we determined the $C$-field as
\begin{align}
C 
\ &= \ 
\frac{1}{2!} C_{ijw} \, \d y^i \wedge \d y^j \wedge \d w
\, .
\label{eq:lKK6_C}
\end{align}
They are used to determine the scaling factors in the other sectors.
By substituting 
\eqref{eq:lKK6_metric} and \eqref{eq:lKK6_C}
into the other parts in the generalized metric \eqref{eq:rescaled_gen_metric} and 
comparing it with \ref{eq:gm_mathematica}, we determine all the scaling factors \eqref{eq:scale3} consistently.

%%%%%%%%%%%%%%%%%%%%%%%%%%%%%%%%%%%%%%%%%%%%%%%%%%%%%%%%%%%%%%%%%%%%%%
%%%%%%%%%%%%%%%%%%%%%%%%%%%%%%%%%%%%%%%%%%%%%%%%%%%%%%%%%%%%%%%%%%%%%%

\section{Superfields in $(2+1)$ dimensions}
\label{sect:SF}

In this appendix we briefly discuss various tools and expansion of superfields in $(2+1)$ dimensions.
Details of the exhibition might be a bit different from literature, though their physical meaning is the same, of course.

%%%%%%%%%%%%%%%%%%%%%%%%%%%%%%%%%
\subsection{Preliminary}

Since Lorentz group in $(2+1)$-dimensional spacetime is 
$SO(2,1) \simeq SL(2,{\mathbb R})$,
the representation ${\bm 2}$ is equal to its conjugate $\ol{\bf 2}$.
Then spinors in this spacetime obey a bit different contraction rule from that in $(3+1)$ dimensions.
That is, we do not have to distinguish dotted spinor indices and undotted ones, for instance, $\ol{\psi}{}_{\dot{\alpha}} = \ol{\psi}{}_{\alpha}$ where $\alpha = 1,2$.
Therefore we define the contraction rule in $(2+1)$ dimensions as follows:
\begin{align}
\theta \psi 
\ &:= \ 
\theta^{\alpha} \psi_{\alpha}
\, , \ls
\ol{\theta} \ol{\psi}
\ := \
\ol{\theta}{}^{\alpha} \ol{\psi}{}_{\alpha}
\ = \ 
- \ol{\theta}{}_{\alpha} \ol{\psi}{}^{\alpha}
\, , 
\end{align}
where we raise and lower the spinor index $\alpha$ in terms of $\ve_{\alpha \beta}$ and $\ve^{\alpha \beta}$ in the same way as \cite{Wess:1992cp}:
$\psi_{\alpha} = \ve_{\alpha \beta} \psi^{\beta}$ and
$\psi^{\alpha} = \ve^{\alpha \beta} \psi_{\beta}$.
Here the explicit form of $\ve_{\alpha \beta}$ is 
$\ve_{12} = -1 = - \ve_{21}$ and $\ve^{12} = +1 = - \ve^{21}$.
Next, we introduce the gamma matrices in $(2+1)$ dimensions. 
It is useful to exhibit the following explicit expression:
\begin{align}
(\gamma^m)_{\alpha \beta}
\ &= \ 
+ (\gamma^m)_{\beta \alpha}
\ = \ 
\big\{
(- {\mathbbm{1}})_{\alpha \beta}
\, , \ 
(\tau_x)_{\alpha \beta}
\, , \ 
(\tau_z)_{\alpha \beta}
\big\}
\ls \text{for \ $m = 0,1,2$}
\, , \label{gamma-def}
\end{align}
where $\tau_x$ and $\tau_z$ are the Pauli matrices\footnote{The spinor indices are raised and lowered in such a way that
$(\gamma^m)^{\alpha}{}_{\beta} =
\ve^{\alpha \tau} (\gamma^m)_{\tau \beta}$, 
$(\gamma^m)_{\alpha}{}^{\beta} =
\ve^{\beta \kappa} (\gamma^m)_{\alpha \kappa}$, and 
$(\gamma^m)^{\alpha \beta} =
\ve^{\alpha \tau} \ve^{\beta \kappa} (\gamma^m)_{\tau \kappa}$.}.
The contraction with two spinors are described as
\begin{align}
\theta \gamma^m \psi
\ &:= \ 
\theta^{\alpha} (\gamma^m)_{\alpha \beta} \psi^{\beta}
\, , \ls
\ol{\theta} \gamma^m \ol{\psi}
\ = \ 
\ol{\theta}{}^{\alpha} (\gamma^m)_{\alpha \beta} \ol{\psi}{}^{\beta}
\, , \ls
\theta \gamma^m \ol{\psi}
\ = \ 
\theta^{\alpha} (\gamma^m)_{\alpha \beta} \ol{\psi}{}^{\beta}
\, . 
\end{align}
We introduce hermitian conjugate of contraction of fermionic spinors in such a way that
$(\theta \psi)^{\dagger} :=
\ol{\psi}{}_{\alpha} \ol{\theta}{}^{\alpha}
= - \ol{\psi} \ol{\theta}
= - \ol{\theta} \ol{\psi}$.
Then the hermitian conjugate of the contraction with the gamma matrix is 
$(\theta \gamma^m \psi)^{\dagger}
= + \ol{\psi} \gamma^m \ol{\theta}
= - \ol{\theta} \gamma^m \ol{\psi}$, 
where we used the gamma matrices (\ref{gamma-def}) are real and symmetric: $(\gamma^m)^*_{\beta \alpha} = (\gamma^m)_{\alpha \beta}$.
We define the supercovariant derivative $D_{\alpha}$ and its anticommutation relation:
\bsubeq
\begin{gather}
D_{\alpha}
\ := \
\frac{\del}{\del \theta^{\alpha}}
+ \I (\gamma^m \ol{\theta})_{\alpha} \del_m 
\, , \ls
\ol{D}{}_{\alpha}
\ = \ 
- \frac{\del}{\del \ol{\theta}{}^{\alpha}}
- \I (\gamma^m \theta)_{\alpha} \del_m
\, , \\
\{ D_{\alpha} , \ol{D}{}_{\beta} \}
\ = \ 
- 2 \I (\gamma^m)_{\alpha \beta} \del_m
\, .
\end{gather}
\esubeq
We define integrals over Grassmann coordinates $\theta$.
The measures are defined as
\bsubeq
\begin{align}
\d^2 \theta
\ := \
+ \frac{1}{2} \d \theta^1 \d \theta^2 
\, , \ls
\d^2 \ol{\theta}
\ = \ 
- \frac{1}{2} \d \ol{\theta}{}^1 \d \ol{\theta}{}^2
\, , \ls
\d^4 \theta 
\ := \ 
\d^2 \theta \, \d^2 \ol{\theta}
\, . 
\end{align}
Then the integrals are given in the following forms:
\begin{align}
\int \d^2 \theta \, (\theta \theta)
\ &= \ 
+1
\, , \ls
\int \d^2 \ol{\theta} \, (\ol{\theta} \ol{\theta})
\ = \ 
-1
\, , \ls
\int \d^4 \theta \, (\theta \theta \ol{\theta} \ol{\theta})
\ = \ 
-1
\, . 
\end{align}
\esubeq

%%%%%%%%%%%%%%%%%%%%%%%%%%%%%%%%%
\subsection{Superfields}
\label{subsect:SF}

In the main text, we use various superfields to construct a gauge theory.
It is worth mentioning their expansion in terms of ordinary fields.%\footnote{Here we omit the $U(1)^N$ gauge group label $\a$.}.

%%%%%%%%%%%%%%%%%%%%%%
%\begin{itemize}
%\item 
{\bf Chiral superfields:}
Chiral superfields $\Phi$, $Q$, $\wt{Q}$, $\Psi$ and $\Gamma$ in the main text are, as in $(3+1)$ dimensions, defined as $0 = \ol{D}{}_{\alpha} \Phi$, and so forth.
Their expansions are given as
\bsubeq
\begin{align}
\Phi
\ &= \ 
\phi
+ \I \sqrt{2} \, (\theta \wt{\lambda})
+ \I (\theta \theta) \, {\sf F}_{\Phi}
\nn \\
\ & \ \ \ \ 
+ \I (\theta \gamma^m \ol{\theta}) \, \del_m \phi
- \frac{1}{\sqrt{2}} (\theta \theta) (\ol{\theta} \gamma^m \del_m \wt{\lambda})
- \frac{1}{4} (\theta \theta \ol{\theta} \ol{\theta}) \, \del_m^2 \phi
%% \, , \\
%% %%%%%
%% \ol{\Phi}{}_{\a}
%% \ &= \ 
%% \ol{\phi}{}_{\a}
%% + \I \sqrt{2} \, (\ol{\theta} \ol{\wt{\lambda}}{}_{\a})
%% + \I (\ol{\theta} \ol{\theta}) \, \ol{\sf F}{}_{\Phi,\a}
%% \nn \\
%% \ & \ \ \ \ 
%% - \I (\theta \gamma^m \ol{\theta}) \, \del_m \ol{\phi}{}_{\a}
%% - \frac{1}{\sqrt{2}} (\ol{\theta} \ol{\theta}) (\theta \gamma^m \del_m \ol{\wt{\lambda}}{}_{\a})
%% - \frac{1}{4} (\theta \theta \ol{\theta} \ol{\theta}) \, \del_m^2 \ol{\phi}{}_{\a}
\, , \label{def-chiral} \\
%\end{align}
%%%%%%%%%%%%%%
%\bsubeq
%\begin{align}
Q
\ &= \ 
q
+ \I \sqrt{2} \, (\theta \psi)
+ \I (\theta \theta) \, {\sf F}
\nn \\
\ & \ \ \ \ 
+ \I (\theta \gamma^m \ol{\theta}) \, \del_m q
- \frac{1}{\sqrt{2}} (\theta \theta) (\ol{\theta} \gamma^m \del_m \psi)
- \frac{1}{4} (\theta \theta \ol{\theta} \ol{\theta}) \, \del_m^2 q
\, , \\
%%%%%
%% \ol{Q}{}_{\a}
%% \ &= \ 
%% \ol{q}{}_{\a}
%% + \I \sqrt{2} \, (\ol{\theta} \ol{\psi}{}_{\a})
%% + \I (\ol{\theta} \ol{\theta}) \, \ol{\sf F}{}_{\a}
%% \nn \\
%% \ & \ \ \ \ 
%% - \I (\theta \gamma^m \ol{\theta}) \, \del_m \ol{q}{}_{\a}
%% - \frac{1}{\sqrt{2}} (\ol{\theta} \ol{\theta}) (\theta \gamma^m \del_m \ol{\psi}{}_{\a})
%% - \frac{1}{4} (\theta \theta \ol{\theta} \ol{\theta}) \, \del_m^2 \ol{q}{}_{\a}
%% \, , \\
%%%%%
\wt{Q}
\ &= \ 
\wt{q}
+ \I \sqrt{2} \, (\theta \wt{\psi})
+ \I (\theta \theta) \, \wt{\sf F}
\nn \\
\ & \ \ \ \ 
+ \I (\theta \gamma^m \ol{\theta}) \, \del_m \wt{q}
- \frac{1}{\sqrt{2}} (\theta \theta) (\ol{\theta} \gamma^m \del_m \wt{\psi})
- \frac{1}{4} (\theta \theta \ol{\theta} \ol{\theta}) \, \del_m^2 \wt{q}
%% \, , \\
%% %%%%%
%% \ol{\wt{Q}}{}_{\a}
%% \ &= \ 
%% \ol{\wt{q}}{}_{\a}
%% + \I \sqrt{2} \, (\ol{\theta} \ol{\wt{\psi}}{}_{\a})
%% + \I (\ol{\theta} \ol{\theta}) \, \ol{\wt{\sf F}}{}_{\a}
%% \nn \\
%% \ & \ \ \ \ 
%% - \I (\theta \gamma^m \ol{\theta}) \, \del_m \ol{\wt{q}}{}_{\a}
%% - \frac{1}{\sqrt{2}} (\ol{\theta} \ol{\theta}) (\theta \gamma^m \del_m \ol{\wt{\psi}}{}_{\a})
%% - \frac{1}{4} (\theta \theta \ol{\theta} \ol{\theta}) \, \del_m^2 \ol{\wt{q}}{}_{\a}
\, , \\
%\end{align}
%\esubeq
%%%%%%%%%%%%%%
%\bsubeq
%\begin{align}
\Psi
\ &= \ 
\frac{1}{\sqrt{2}} (r^1 + \I r^2)
+ \I \sqrt{2} \, (\theta \chi)
+ \I (\theta \theta) \, {\sf G}_{\Psi}
\nn \\
\ & \ \ \ \ 
+ \frac{\I}{\sqrt{2}} (\theta \gamma^m \ol{\theta}) \, \del_m (r^1 + \I r^2)
- \frac{1}{\sqrt{2}} (\theta \theta) (\ol{\theta} \gamma^m \del_m \chi)
- \frac{1}{4 \sqrt{2}} (\theta \theta \ol{\theta} \ol{\theta}) \, \del_m^2 (r^1 + \I r^2)
\, , \\
%%%%%
%% \ol{\Psi}
%% \ &= \ 
%% \frac{1}{\sqrt{2}} (r^1 - \I r^2)
%% + \I \sqrt{2} \, (\ol{\theta} \ol{\chi})
%% + \I (\ol{\theta} \ol{\theta}) \, \ol{\sf G}{}_{\Psi}
%% \nn \\
%% \ & \ \ \ \ 
%% - \frac{\I}{\sqrt{2}} (\theta \gamma^m \ol{\theta}) \, \del_m (r^1 - \I r^2)
%% - \frac{1}{\sqrt{2}} (\ol{\theta} \ol{\theta}) (\theta \gamma^m \del_m \ol{\chi})
%% - \frac{1}{4 \sqrt{2}} (\theta \theta \ol{\theta} \ol{\theta}) \, \del_m^2 (r^1 - \I r^2)
%% \, , \\
%%%%%
\Gamma
\ &= \ 
\frac{1}{\sqrt{2}} (\wt{r}^3 + \I \wt{r}^4)
+ \I \sqrt{2} \, (\theta \wt{\chi})
+ \I (\theta \theta) \, {\sf G}_{\Gamma}
\nn \\
\ & \ \ \ \ 
+ \frac{\I}{\sqrt{2}} (\theta \gamma^m \ol{\theta}) \, \del_m (\wt{r}^3 + \I \wt{r}^4)
- \frac{1}{\sqrt{2}} (\theta \theta) (\ol{\theta} \gamma^m \del_m \wt{\chi})
- \frac{1}{4 \sqrt{2}} (\theta \theta \ol{\theta} \ol{\theta}) \, \del_m^2 (\wt{r}^3 + \I \wt{r}^4)
%% \, , \\
%% %%%%%
%% \ol{\Gamma}
%% \ &= \ 
%% \frac{1}{\sqrt{2}} (\wt{r}^3 - \I \wt{r}^4)
%% + \I \sqrt{2} \, (\ol{\theta} \ol{\wt{\chi}})
%% + \I (\ol{\theta} \ol{\theta}) \, \ol{\sf G}{}_{\Gamma}
%% \nn \\
%% \ & \ \ \ \ 
%% - \frac{\I}{\sqrt{2}} (\theta \gamma^m \ol{\theta}) \, \del_m (\wt{r}^3 - \I \wt{r}^4)
%% - \frac{1}{\sqrt{2}} (\ol{\theta} \ol{\theta}) (\theta \gamma^m \del_m \ol{\wt{\chi}})
%% - \frac{1}{4 \sqrt{2}} (\theta \theta \ol{\theta} \ol{\theta}) \, \del_m^2 (\wt{r}^3 - \I \wt{r}^4)
\, . 
\end{align}
\esubeq

%%%%%%%%%%%%%%%%%%%%%%
%\item 
{\bf Vector superfield in the Wess-Zumino gauge:}
A vector superfield $V$ also appears in $(3+1)$ dimensions. This is
defined by the reality condition $V^{\dagger} = V$.
Removing many redundant degrees of freedom via the Wess-Zumino gauge fixing, we describe $V_{\text{WZ}}$ as
\begin{align}
V_{\text{WZ}}
\ &= \ 
- \I (\theta \ol{\theta}) \, \sigma
- (\theta \gamma^m \ol{\theta}) \, A_m
- \I (\theta \theta) (\ol{\theta} \ol{\lambda})
+ \I (\ol{\theta} \ol{\theta}) (\theta \lambda)
- \half (\theta \theta \ol{\theta} \ol{\theta}) \, {\sf D}
\, . \label{def-vector}
\end{align}

%%%%%%%%%%%%%%%%%%%%%%
%\item 
{\bf Real linear superfield for gauge field strength:}
A real linear superfield $\Sigma = \Sigma^{\dagger}$ is generically defined as
$0 = \ol{D}{}^{\alpha} \ol{D}{}_{\alpha} \Sigma = D^{\alpha} D_{\alpha} \Sigma$.
If we set a real linear superfield to $\Sigma := 
+ \textstyle{\frac{\I}{2}} \ol{D}{}^{\alpha} D_{\alpha} V_{\text{WZ}}$,
this contains an (Abelian) gauge field strength $F_{mn} = \del_m A_{n} - \del_n A_m$ in such a way that\footnote{This is analogous to a spinorial superfield $W_{\alpha} = - \frac{1}{4} \ol{D} \ol{D} D_{\alpha} V$ in \cite{Wess:1992cp}.}
%\bsubeq 
\begin{align}
\Sigma
\ &= \
\sigma
- (\theta \ol{\lambda})
+ (\ol{\theta} \lambda)
+ \I (\theta \ol{\theta}) \, {\sf D}
- \half (\theta \gamma_p \ol{\theta}) \, {\ve}^{mnp} {F}_{mn}
\nn \vf \\
\ & \ \ \ \ 
+ \frac{\I}{2} (\theta \theta) (\ol{\theta} \gamma^m \del_m \ol{\lambda})
- \frac{\I}{2} (\ol{\theta} \ol{\theta}) (\theta \gamma^m \del_m \lambda)
+ \frac{1}{4} (\theta \theta \ol{\theta} \ol{\theta}) \, \del_m^2 \sigma
\, , \label{def-reallinear}
\end{align}
where ${\ve}^{mnp}$ is the totally antisymmetric symbol in $(2+1)$ dimensions whose normalization is ${\ve}^{012} = +1$.

\end{appendix}
%%%%%%%%%%%%%%%%%%%%%%%%%%%%%%%%%%%%%%%%%%%%%%%%%%%%%%%%%%%%%%%%%%%%%%
%%%%%%%%%%%%%%%%%%%%%%%%%%%%%%%%%%%%%%%%%%%%%%%%%%%%%%%%%%%%%%%%%%%%%%
%\newpage
\phantomsection
\addcontentsline{toc}{section}{References}
{

}
%%%%%%%%%%%%%%%%%%%%%%%%%%%%%%%%%%%%%%%%%%%%%%%%%%%%%%%%%%%%%%%%%%%%%%
%%%%%%%%%%%%%%%%%%%%%%%%%%%%%%%%%%%%%%%%%%%%%%%%%%%%%%%%%%%%%%%%%%%%%%
}

\begin{thebibliography}{99}

\bibitem{Gregory:1997te}
  R.~Gregory, J.~A.~Harvey and G.~W.~Moore,
  ``{Unwinding Strings and T-duality of Kaluza-Klein and H-monopoles},''
  Adv.\ Theor.\ Math.\ Phys.\  {\bf 1} (1997) 283
%  doi:10.4310/ATMP.1997.v1.n2.a6
  [hep-th/9708086].
  %%CITATION = doi:10.4310/ATMP.1997.v1.n2.a6;%%


\bibitem{Buscher:1987sk}
T.~H.~Buscher,
``{A Symmetry of the String Background Field Equations},''
Phys. Lett. B \textbf{194} (1987), 59-62, 
%doi:10.1016/0370-2693(87)90769-6
%898 citations counted in INSPIRE as of 14 Jul 2021
%
%\cite{Buscher:1987qj}
%\bibitem{Buscher:1987qj}
%T.~H.~Buscher,
``{Path Integral Derivation of Quantum Duality in Nonlinear Sigma Models},''
Phys. Lett. B \textbf{201} (1988), 466-472.
%doi:10.1016/0370-2693(88)90602-8
%918 citations counted in INSPIRE as of 14 Jul 2021

\bibitem{Wen:1985jz}
  X.~G.~Wen and E.~Witten,
  ``{World-sheet Instantons and the {Peccei-Quinn} Symmetry},''
  Phys.\ Lett.\  {\bf 166B} (1986) 397.
  %doi:10.1016/0370-2693(86)91587-X
  %%CITATION = doi:10.1016/0370-2693(86)91587-X;%%
  %131 citations counted in INSPIRE as of 04 Jan 2018


\bibitem{Tong:2002rq}
  D.~Tong,
  ``{NS5-branes, T-duality and Worldsheet Instantons},''
  JHEP {\bf 0207} (2002) 013
%  doi:10.1088/1126-6708/2002/07/013
  [hep-th/0204186].
  %%CITATION = doi:10.1088/1126-6708/2002/07/013;%%

\bibitem{Harvey:2005ab}
  J.~A.~Harvey and S.~Jensen,
  ``{Worldsheet Instanton Corrections to the Kaluza-Klein Monopole},''
  JHEP {\bf 0510} (2005) 028
%  doi:10.1088/1126-6708/2005/10/028
  [hep-th/0507204].
  %%CITATION = doi:10.1088/1126-6708/2005/10/028;%%



\bibitem{Jensen:2011jna}
  S.~Jensen,
  ``{The KK-monopole/NS5-brane in Doubled Geometry},''
  JHEP {\bf 1107} (2011) 088
  %doi:10.1007/JHEP07(2011)088
  [arXiv:1106.1174 [hep-th]].
  %%CITATION = doi:10.1007/JHEP07(2011)088;%%
  %30 citations counted in INSPIRE as of 21 Nov 2017


\bibitem{Kimura:2013zva}
  T.~Kimura and S.~Sasaki,
  ``{Worldsheet Instanton Corrections to $5^2_2$-brane Geometry},''
  JHEP {\bf 1308} (2013) 126
  %doi:10.1007/JHEP08(2013)126
  [arXiv:1305.4439 [hep-th]].
  %%CITATION = doi:10.1007/JHEP08(2013)126;%%
  %30 citations counted in INSPIRE as of 21 Nov 2017


\bibitem{Hull:2009mi}
  C.~Hull and B.~Zwiebach,
  ``{Double Field Theory},''
  JHEP {\bf 0909} (2009) 099
  %doi:10.1088/1126-6708/2009/09/099
  [arXiv:0904.4664 [hep-th]].
  %%CITATION = doi:10.1088/1126-6708/2009/09/099;%%
  %375 citations counted in INSPIRE as of 21 Nov 2017


\bibitem{Siegel:1993xq}
  W.~Siegel,
  ``{Two Vierbein Formalism for String Inspired Axionic Gravity},''
  Phys.\ Rev.\ D {\bf 47} (1993) 5453
  %doi:10.1103/PhysRevD.47.5453
  [hep-th/9302036],
  %%CITATION = doi:10.1103/PhysRevD.47.5453;%%
  %284 citations counted in INSPIRE as of 24 May 2018
%\bibitem{Siegel:1993th}
  %W.~Siegel,
  ``{Superspace Duality in Low-energy Superstrings},''
  Phys.\ Rev.\ D {\bf 48} (1993) 2826
  %doi:10.1103/PhysRevD.48.2826
  [hep-th/9305073],
  %%CITATION = doi:10.1103/PhysRevD.48.2826;%%
  %342 citations counted in INSPIRE as of 24 May 2018
%\bibitem{Siegel:1993bj}
  %W.~Siegel,
  ``{Manifest Duality in Low-energy Superstrings},''
  hep-th/9308133.
  %%CITATION = HEP-TH/9308133;%%
  %104 citations counted in INSPIRE as of 24 May 2018


\bibitem{Berman:2014jsa}
  D.~S.~Berman and F.~J.~Rudolph,
  ``{Branes are Waves and Monopoles},''
  JHEP {\bf 1505} (2015) 015
  %doi:10.1007/JHEP05(2015)015
  [arXiv:1409.6314 [hep-th]].
  %%CITATION = doi:10.1007/JHEP05(2015)015;%%
  %38 citations counted in INSPIRE as of 21 Nov 2017

\bibitem{Bakhmatov:2016kfn}
  I.~Bakhmatov, A.~Kleinschmidt and E.~T.~Musaev,
  ``{Non-geometric Branes are DFT Monopoles},''
  JHEP {\bf 1610} (2016) 076
  %doi:10.1007/JHEP10(2016)076
  [arXiv:1607.05450 [hep-th]].
  %%CITATION = doi:10.1007/JHEP10(2016)076;%%
  %12 citations counted in INSPIRE as of 21 Nov 2017


\bibitem{Kimura:2018hph}
T.~Kimura, S.~Sasaki and K.~Shiozawa,
``{Worldsheet Instanton Corrections to Five-branes and Waves in Double Field Theory},''
JHEP \textbf{07} (2018), 001
%doi:10.1007/JHEP07(2018)001
[arXiv:1803.11087 [hep-th]].
%17 citations counted in INSPIRE as of 23 Jun 2021




\bibitem{Hohm:2013pua}
O.~Hohm and H.~Samtleben,
``{Exceptional Form of $D=11$ Supergravity},''
Phys. Rev. Lett. \textbf{111} (2013), 231601
%doi:10.1103/PhysRevLett.111.231601
[arXiv:1308.1673 [hep-th]].
%177 citations counted in INSPIRE as of 10 Jul 2021
%
\bibitem{Hohm:2013vpa}
O.~Hohm and H.~Samtleben,
``{Exceptional Field Theory I: $E_{6(6)}$ covariant Form of M-Theory and Type IIB},''
Phys. Rev. D \textbf{89} (2014) no.6, 066016
%doi:10.1103/PhysRevD.89.066016
[arXiv:1312.0614 [hep-th]], 
%200 citations counted in INSPIRE as of 10 Jul 2021
%
%\bibitem{Hohm:2013uia}
%O.~Hohm and H.~Samtleben,
``{Exceptional Field Theory. II. $E_{7(7)}$},''
Phys. Rev. D \textbf{89} (2014), 066017
%doi:10.1103/PhysRevD.89.066017
[arXiv:1312.4542 [hep-th]],
%202 citations counted in INSPIRE as of 10 Jul 2021
%
%\bibitem{Hohm:2014fxa}
%O.~Hohm and H.~Samtleben,
``{Exceptional Field Theory. III. $E_{8(8)}$},''
Phys. Rev. D \textbf{90} (2014), 066002
%doi:10.1103/PhysRevD.90.066002
[arXiv:1406.3348 [hep-th]].
%147 citations counted in INSPIRE as of 10 Jul 2021


\bibitem{Abzalov:2015ega}
A.~Abzalov, I.~Bakhmatov and E.~T.~Musaev,
``{Exceptional Field Theory: $SO(5,5)$},''
JHEP \textbf{06} (2015), 088
%doi:10.1007/JHEP06(2015)088
[arXiv:1504.01523 [hep-th]].
%65 citations counted in INSPIRE as of 10 Jul 2021


\bibitem{Musaev:2015ces}
E.~T.~Musaev,
``{Exceptional Field Theory: $SL(5)$},''
JHEP \textbf{02} (2016), 012
%doi:10.1007/JHEP02(2016)012
[arXiv:1512.02163 [hep-th]].
%62 citations counted in INSPIRE as of 10 Jul 2021


\bibitem{Berman:2015rcc}
D.~S.~Berman, C.~D.~A.~Blair, E.~Malek and F.~J.~Rudolph,
``{An Action for F-theory: ${SL}(2) \times {{\mathbb{R}}}^{+}$ Exceptional Field Theory},''
Class. Quant. Grav. \textbf{33} (2016) no.19, 195009
%doi:10.1088/0264-9381/33/19/195009
[arXiv:1512.06115 [hep-th]].
%53 citations counted in INSPIRE as of 10 Jul 2021


\bibitem{Bossard:2018utw}
G.~Bossard, F.~Ciceri, G.~Inverso, A.~Kleinschmidt and H.~Samtleben,
``{$E_{9}$ Exceptional Field Theory. Part I. The Potential},''
JHEP \textbf{03} (2019), 089
%doi:10.1007/JHEP03(2019)089
[arXiv:1811.04088 [hep-th]], 
%18 citations counted in INSPIRE as of 10 Jul 2021
%\bibitem{Bossard:2021jix}
%G.~Bossard, F.~Ciceri, G.~Inverso, A.~Kleinschmidt and H.~Samtleben,
``{$E_{9}$ Exceptional Field Theory. Part II. The Complete Dynamics},''
JHEP \textbf{05} (2021), 107
%doi:10.1007/JHEP05(2021)107
[arXiv:2103.12118 [hep-th]].
%1 citations counted in INSPIRE as of 10 Jul 2021


\bibitem{Bossard:2019ksx}
G.~Bossard, A.~Kleinschmidt and E.~Sezgin,
``{On Supersymmetric $E_{11}$ Exceptional Field Theory},''
JHEP \textbf{10} (2019), 165
%doi:10.1007/JHEP10(2019)165
[arXiv:1907.02080 [hep-th]].
%13 citations counted in INSPIRE as of 14 Jul 2021


\bibitem{Berman:2020tqn}
D.~S.~Berman and C.~D.~A.~Blair,
``{The Geometry, Branes and Applications of Exceptional Field Theory},''
Int. J. Mod. Phys. A \textbf{35} (2020) no.30, 2030014
%doi:10.1142/S0217751X20300148
[arXiv:2006.09777 [hep-th]].
%29 citations counted in INSPIRE as of 10 Jul 2021




\bibitem{Hohm:2010pp}
  O.~Hohm, C.~Hull and B.~Zwiebach,
  ``{Generalized Metric Formulation of Double Field Theory},''
  JHEP {\bf 1008} (2010) 008
  %doi:10.1007/JHEP08(2010)008
  [arXiv:1006.4823 [hep-th]].
  %%CITATION = doi:10.1007/JHEP08(2010)008;%%
  %288 citations counted in INSPIRE as of 27 Nov 2017


\bibitem{Berkeley:2014nza}
  J.~Berkeley, D.~S.~Berman and F.~J.~Rudolph,
  ``{Strings and Branes are Waves},''
  JHEP {\bf 1406} (2014) 006
  %doi:10.1007/JHEP06(2014)006
  [arXiv:1403.7198 [hep-th]].
  %%CITATION = doi:10.1007/JHEP06(2014)006;%%
  %37 citations counted in INSPIRE as of 21 Nov 2017

\bibitem{Blair:2016xnn}
  C.~D.~A.~Blair,
  ``{Doubled strings, Negative Strings and Null Waves},''
  JHEP {\bf 1611} (2016) 042
  %doi:10.1007/JHEP11(2016)042
  [arXiv:1608.06818 [hep-th]].
  %%CITATION = doi:10.1007/JHEP11(2016)042;%%
  %4 citations counted in INSPIRE as of 21 Nov 2017


\bibitem{Lust:2017jox}
  D.~L\"{u}st, E.~Plauschinn and V.~Vall Camell,
  ``{Unwinding Strings in Semi-flatland},''
  JHEP {\bf 1707} (2017) 027
  %doi:10.1007/JHEP07(2017)027
  [arXiv:1706.00835 [hep-th]].
  %%CITATION = doi:10.1007/JHEP07(2017)027;%%
  %1 citations counted in INSPIRE as of 21 Nov 2017


\bibitem{Berman:2018okd}
D.~S.~Berman, E.~T.~Musaev and R.~Otsuki,
``{Exotic Branes in Exceptional Field Theory: $E_{7(7)}$ and Beyond},''
JHEP \textbf{12} (2018), 053
%doi:10.1007/JHEP12(2018)053
[arXiv:1806.00430 [hep-th]].
%22 citations counted in INSPIRE as of 17 Jul 2021

\bibitem{Berman:2011jh}
D.~S.~Berman, H.~Godazgar, M.~J.~Perry and P.~West,
``{Duality Invariant Actions and Generalised Geometry},''
JHEP \textbf{02} (2012), 108
%doi:10.1007/JHEP02(2012)108
[arXiv:1111.0459 [hep-th]].
%162 citations counted in INSPIRE as of 10 Jul 2021

%\cite{Coimbra:2011ky}
\bibitem{Coimbra:2011ky}
A.~Coimbra, C.~Strickland-Constable and D.~Waldram,
``{$E_{d(d)} \times \mathbb{R}^+$ Generalised Geometry, Connections and M-theory},''
JHEP \textbf{02} (2014), 054
%doi:10.1007/JHEP02(2014)054
[arXiv:1112.3989 [hep-th]].
%224 citations counted in INSPIRE as of 27 Aug 2021

\bibitem{Okuyama:2005gx}
  K.~Okuyama,
  ``{Linear Sigma Models of H and KK Monopoles},''
  JHEP {\bf 0508} (2005) 089
%  doi:10.1088/1126-6708/2005/08/089
  [hep-th/0508097].
  %%CITATION = doi:10.1088/1126-6708/2005/08/089;%%

\bibitem{Becker:1995kb}
K.~Becker, M.~Becker and A.~Strominger,
``{Five-branes, Membranes and Nonperturbative String Theory},''
Nucl. Phys. B \textbf{456} (1995), 130-152
%doi:10.1016/0550-3213(95)00487-1
[arXiv:hep-th/9507158 [hep-th]].
%727 citations counted in INSPIRE as of 26 May 2021

\bibitem{Harvey:1999as}
J.~A.~Harvey and G.~W.~Moore,
``{Superpotentials and Membrane Instantons},''
[arXiv:hep-th/9907026 [hep-th]].
%203 citations counted in INSPIRE as of 17 Jul 2021

\bibitem{Kapustin:1999ha}
  A.~Kapustin and M.~J.~Strassler,
  ``{On Mirror Symmetry in Three-dimensional Abelian Gauge Theories},''
  JHEP {\bf 9904} (1999) 021
%  doi:10.1088/1126-6708/1999/04/021
  [hep-th/9902033].
  %%CITATION = doi:10.1088/1126-6708/1999/04/021;%%

\bibitem{Kimura:2013fda}
T.~Kimura and S.~Sasaki,
``{Gauged Linear Sigma Model for Exotic Five-brane},''
Nucl. Phys. B \textbf{876} (2013), 493-508
%doi:10.1016/j.nuclphysb.2013.08.017
[arXiv:1304.4061 [hep-th]].
%53 citations counted in INSPIRE as of 29 Jun 2021

\bibitem{Intriligator:1996ex}
K.~A.~Intriligator and N.~Seiberg,
``{Mirror Symmetry in Three-dimensional Gauge Theories},''
Phys. Lett. B \textbf{387} (1996), 513-519
%doi:10.1016/0370-2693(96)01088-X
[arXiv:hep-th/9607207 [hep-th]].
%552 citations counted in INSPIRE as of 01 Sep 2021

\bibitem{Kapustin:2010xq}
A.~Kapustin, B.~Willett and I.~Yaakov,
``{Nonperturbative Tests of Three-dimensional Dualities},''
JHEP \textbf{10} (2010), 013
%doi:10.1007/JHEP10(2010)013
[arXiv:1003.5694 [hep-th]].
%207 citations counted in INSPIRE as of 01 Sep 2021


\bibitem{deBoer:1996mp}
J.~de Boer, K.~Hori, H.~Ooguri and Y.~Oz,
``{Mirror Symmetry in Three-dimensional Gauge theories, Quivers and D-branes},''
Nucl. Phys. B \textbf{493} (1997), 101-147
%doi:10.1016/S0550-3213(97)00125-9
[arXiv:hep-th/9611063 [hep-th]].
%207 citations counted in INSPIRE as of 01 Sep 2021

\bibitem{Hanany:1996ie}
A.~Hanany and E.~Witten,
``{Type IIB Superstrings, BPS Monopoles, and Three-dimensional Gauge Dynamics},''
Nucl. Phys. B \textbf{492} (1997), 152-190
%doi:10.1016/S0550-3213(97)00157-0
[arXiv:hep-th/9611230 [hep-th]].
%1210 citations counted in INSPIRE as of 01 Sep 2021

\bibitem{Bullimore:2015lsa}
M.~Bullimore, T.~Dimofte and D.~Gaiotto,
``{The Coulomb Branch of 3d ${\mathcal{N}= 4}$ Theories},''
Commun. Math. Phys. \textbf{354} (2017) no.2, 671-751
%doi:10.1007/s00220-017-2903-0
[arXiv:1503.04817 [hep-th]].
%125 citations counted in INSPIRE as of 01 Sep 2021

\bibitem{Brooks:1994nn}
R.~Brooks and S.~J.~Gates, Jr.,
``{Extended Supersymmetry and Super-BF Gauge Theories},''
Nucl. Phys. B \textbf{432} (1994), 205-224
%doi:10.1016/0550-3213(94)90600-9
[arXiv:hep-th/9407147 [hep-th]].
%37 citations counted in INSPIRE as of 16 Jul 2021

%\bibitem{toappear}
%T.~Kimura, S.~Sasaki, K.~Shiozawa, work in progress.

\bibitem{Harvey:2014nha}
J.~A.~Harvey, S.~Lee and S.~Murthy,
``{Elliptic Genera of ALE and ALF Manifolds from Gauged Linear Sigma Models},''
JHEP \textbf{02} (2015), 110
%doi:10.1007/JHEP02(2015)110
[arXiv:1406.6342 [hep-th]].
%43 citations counted in INSPIRE as of 16 Jul 2021


\bibitem{Hiraga:2020lhk}
Y.~Hiraga and Y.~Sato,
``{Localization of the Gauged Linear Sigma Model for KK5-branes},''
PTEP \textbf{2021} (2021) no.3, 033B06
%doi:10.1093/ptep/ptab031
[arXiv:2011.06919 [hep-th]].
%0 citations counted in INSPIRE as of 16 Jul 2021

\bibitem{Wess:1992cp}
J.~Wess and J.~Bagger,
``{Supersymmetry and Supergravity, 2nd Edition},''
Princeton University Press.

\end{thebibliography}
\end{document}